\pdfoutput=1
\documentclass[11pt]{amsart}
\usepackage{multirow}
\usepackage{graphicx}
\usepackage{amsmath}
\usepackage{amssymb}
\usepackage{textcomp}
\usepackage[margin=1in]{geometry}
\usepackage[square,numbers,sort]{natbib}
\usepackage[colorlinks=true,linkcolor=black,citecolor=true,
citecolor=black,breaklinks=true,
pdfpagemode=UseThumbs,bookmarksnumbered=true]{hyperref}

\allowdisplaybreaks

\graphicspath{{figs/wct/examples/},{figs/wct/meshes/},{figs/wct/misc/}}

\newcommand{\RR}{\ensuremath{\mathbb{R}}}
\newcommand{\ZZ}{\ensuremath{\mathbb{Z}}}

\begin{document}
\title[Triangulation of Simple 3D Shapes with Well-Centered Tetrahedra]
{Triangulation of Simple 3D Shapes \\ with Well-Centered Tetrahedra}

\author{Evan VanderZee} 

\address{Department of Mathematics, 1409 W. Green Street, University
  of Illinois at Urbana-Champaign, Urbana, IL 61801 }

\email{vanderze@illinois.edu} 

\author{Anil N. Hirani}

\address{Correspondence to: Professor Anil N. Hirani, Department of
  Computer Science, University of Illinois at Urbana-Champaign, 201
  N. Goodwin Ave., Urbana, IL 61801.}

\urladdr{\url{http://www.cs.uiuc.edu/hirani}}

\email{hirani@cs.uiuc.edu}

\author{Damrong Guoy} 

\address{Center for Simulation of Advanced Rockets, University of
  Illinois at Urbana-Champaign, Urbana, IL 61801}

\email{guoy@uiuc.edu}

\begin{abstract}
  A completely well-centered tetrahedral mesh is a triangulation of a
  three dimensional domain in which every tetrahedron and every
  triangle contains its circumcenter in its interior. Such meshes have
  applications in scientific computing and other fields. We show how
  to triangulate simple domains using completely well-centered
  tetrahedra. The domains we consider here are space, infinite slab,
  infinite rectangular prism, cube, and regular tetrahedron. We also
  demonstrate single tetrahedra with various combinations of the
  properties of dihedral acuteness, 2-well-centeredness, and
  3-well-centeredness.

\end{abstract}

\maketitle

\section{Introduction}
\label{sec:intro}

In this paper we demonstrate well-centered triangulation of simple
domains in $\mathbb{R}^3$. A well-centered simplex is one for which
the circumcenter lies in the interior of the simplex
\cite{Hirani2003}.  This definition is further refined to that of a
$k$-well-centered simplex which is one whose $k$-dimensional faces
have the well-centeredness property. An $n$-dimensional simplex which
is $k$-well-centered for all $1 \le k \le n$ is called completely
well-centered \cite{VaHiGuRa2008}. These properties extend to
simplicial complexes, i.e. to meshes. Thus a mesh can be completely
well-centered or $k$-well-centered if all its simplices have that
property. For triangles, being well-centered is the same as being
acute-angled. But a tetrahedron can be dihedral acute without being
3-well-centered as we show by example in Sect.~\ref{sec:single_tet}.
We also note that while every well-centered triangulation is Delaunay,
the converse is not true.

In \cite{VaHiGuRa2007} we described an optimization-based approach to
transform a given planar triangle mesh into a well-centered one by
moving the internal vertices while keeping the boundary vertices
fixed. In \cite{VaHiGuRa2008} we generalized this approach to
arbitrary dimensions and
in addition to developing some theoretical properties of our method
showed some complex examples of our method at
work in the plane and some simple examples in $\mathbb{R}^3$. In
this paper the domains we consider are space, slabs, infinite
rectangular prisms, cubes and tetrahedra.

Well-centered triangulations find applications in some areas of
scientific computing and other fields. Their limited use so far may be
partly because until recently there were no methods known for
constructing such meshes. One motivation for constructing
well-centered meshes comes from Discrete Exterior Calculus, which is a
framework for constructing numerical methods for partial differential
equations~\cite{Hirani2003,DeHiLeMa2005}. The availability of
well-centered meshes permits one to discretize an important operator
called the Hodge star as a diagonal matrix, leading to efficiencies in
numerical solution procedures. Other potential applications are the
covolume method~\cite{Nicolaides1992,SaHaMoWe2006}, space-time
meshing~\cite{UnSh2002}, and computations of geodesic paths on
manifolds~\cite{KiSe1998}.

\section{Well-centeredness and Dihedral Acuteness
  for a Single Tetrahedron}
\label{sec:single_tet}

\begin{figure}
\centering
\includegraphics[width=100pt, trim=388pt 209pt 349pt 176pt, clip]
  {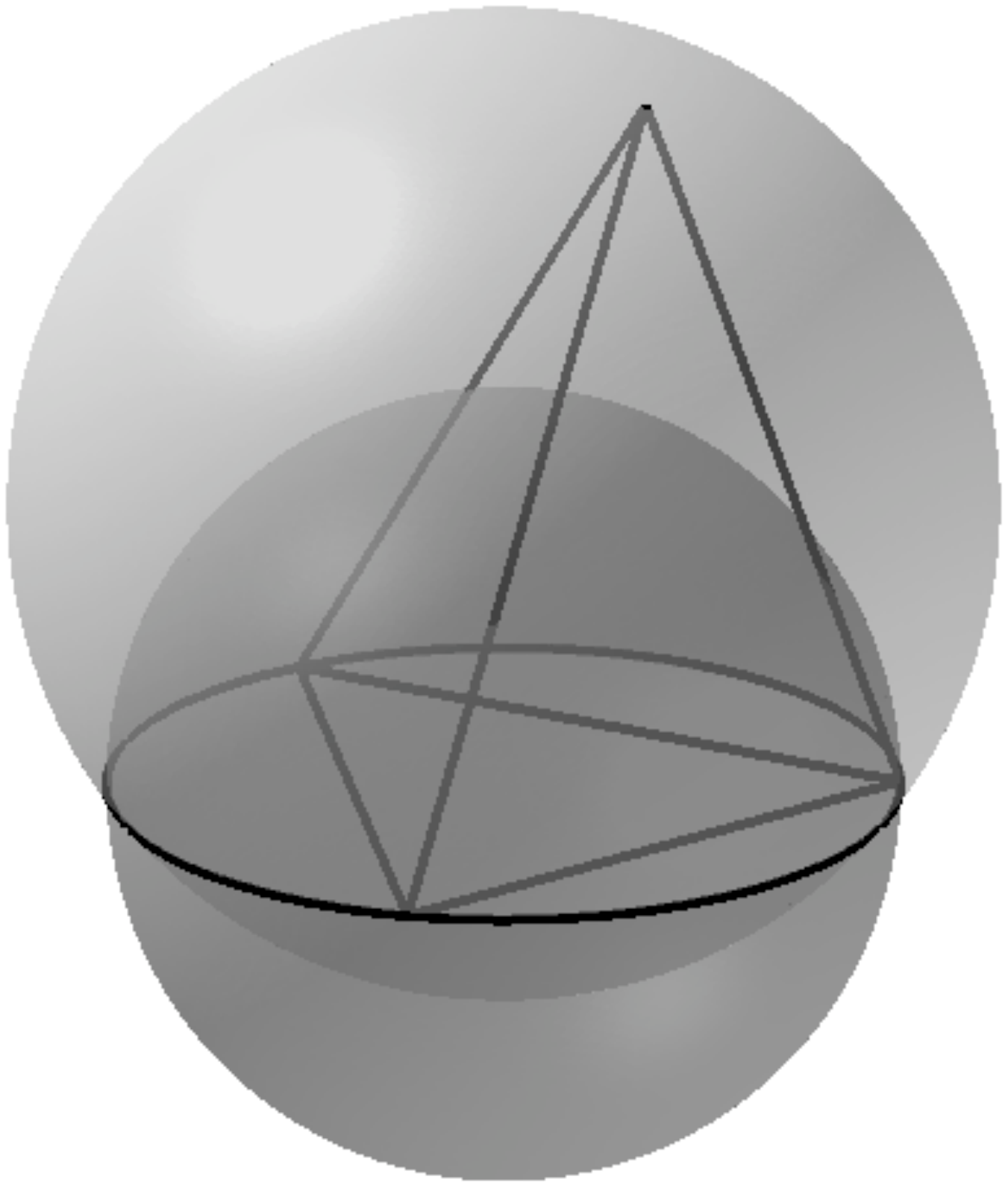}%
\caption{One characterization of well-centeredness is that for each
    vertex $v$, the vertex lies outside of the equatorial ball of the
    facet opposite $v$. The larger sphere shown here is the
    circumsphere of the tetrahedron and the smaller sphere is the
    boundary of the equatorial ball of the bottom triangle. The top
    vertex of the tetrahedron is outside the equatorial ball}
\label{fig:eqball}
\end{figure}

An equatorial ball of a simplex is a ball for which the
circumsphere of the simplex is an equator.  Stated more precisely,
if $\sigma^{k}$ is a $k$-dimensional simplex, then the equatorial
ball of $\sigma^{k}$ is a $(k+1)$-dimensional ball whose center
is $c(\sigma^{k})$, the circumcenter of $\sigma^{k}$, and whose
radius is $R(\sigma^{k})$, the circumradius of $\sigma^{k}$.
In \cite{VaHiGuRa2008} we showed that a simplex $\sigma^{n}$
is $n$-well-centered if and only if for each vertex
$v$ in $\sigma^{n}$, $v$ lies outside the equatorial
ball of the facet $\tau_{v}^{n-1}$ opposite $v$.
Figure~\ref{fig:eqball} illustrates this characterization
of well-centeredness; the vertex $v$ at the top of the
tetrahedron in Fig.~\ref{fig:eqball} lies outside of
the equatorial ball of the facet opposite $v$.

We now have a definition of $k$-well-centeredness and an
alternate characterization of well-cen\-tered\-ness, but these provide
limited intuition for what it means to be well-centered.  In this section we
discuss a variety of tetrahedra, showing that in $\RR^3$, a simplex
that is $2$-well-centered may or may not be $3$-well-centered, and
vice versa.  We also discuss how being dihedral acute relates to being
well-centered.

\begin{figure}[p]
\centering
\begin{minipage}[c]{160pt}
\includegraphics[width=150pt, trim=266pt 266pt 220pt 222pt, clip]
  {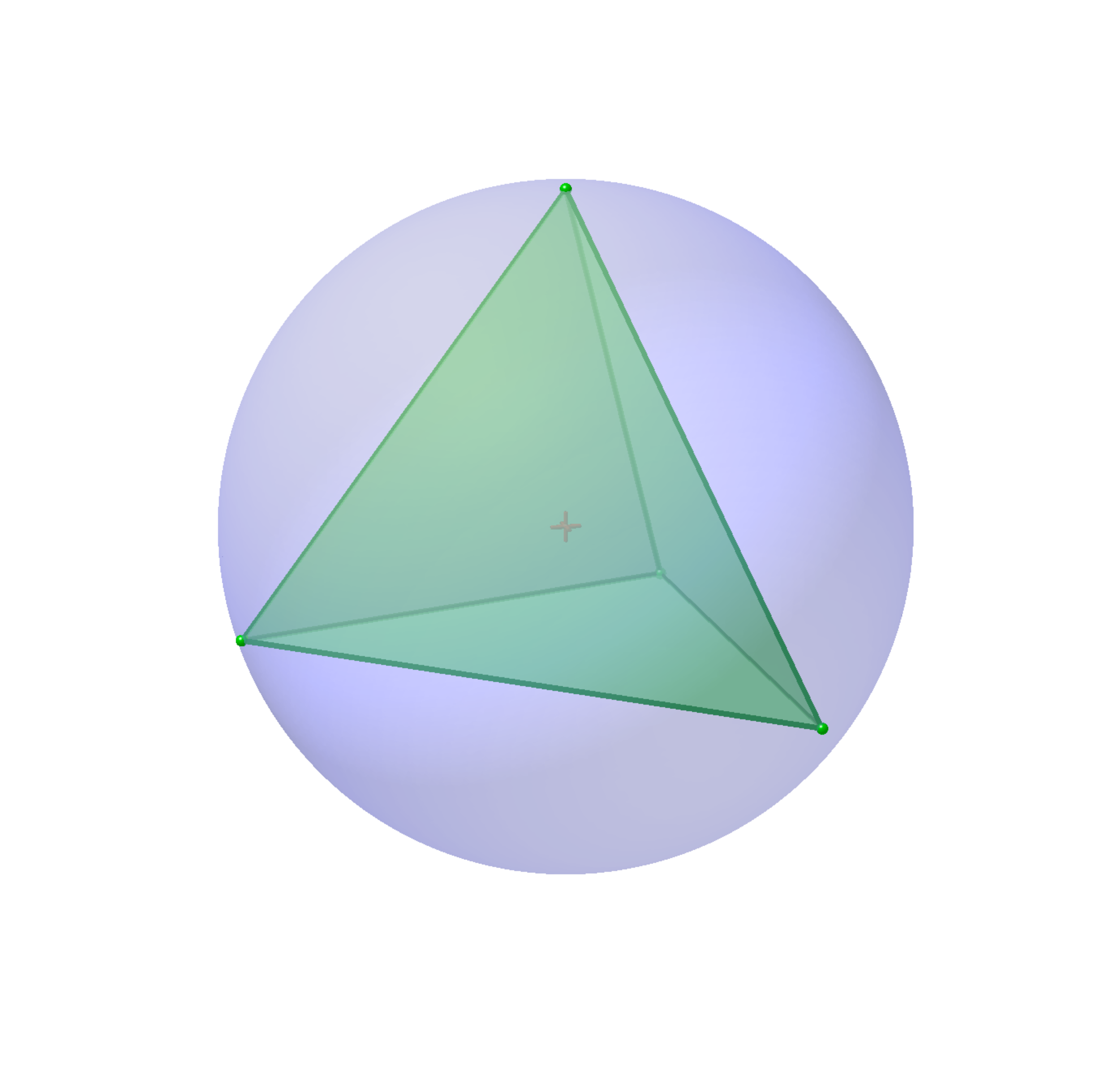}%
\end{minipage}%
\hspace{10pt}
\begin{minipage}[c]{200pt}
\centering
\begin{tabular}{|l|l|l|}
\hline
\multicolumn{3}{|c|}{Vertex Coordinates}\\
\hline
\multicolumn{1}{|c|}{$x$} & \multicolumn{1}{c|}{$y$}
  & \multicolumn{1}{c|}{$z$}\\
\hline
\ \ $\phantom{-}0.6\phantom{000}$ & \ \ $-0.64\phantom{00}$
  & \ \ $-0.48\phantom{00}$\\
\hline
\ \ $\phantom{-}0.48$ & \ \ $\phantom{-}0.8$ & \ \ $-0.36$\\
\hline
\ \ $-0.96$ & \ \ $\phantom{-}0$ & \ \ $-0.28$\\
\hline
\ \ $\phantom{-}0$ & \ \ $\phantom{-}0$ & \ \ $\phantom{-}1$\\
\hline
\end{tabular}\\[10pt]

\begin{tabular}{|l|l|l|}
\hline
\multicolumn{3}{|c|}{Quality Statistics}\\
\hline
\multicolumn{1}{|c|}{Quantity} & \multicolumn{1}{c|}{Min}
  & \multicolumn{1}{c|}{Max}\\
\hline
\ $h/R$ & \ \ $0.254$ & \ \ $0.371$\\
\hline
\ Face Angle & \ \ $50.92$\textdegree\ \  & \ \ $67.08$\textdegree\ \ \\
\hline
\ Dihedral Angle\ \ & \ \ $58.76$\textdegree & \ \ $76.98$\textdegree\\
\hline
\ $R/\ell$ & \ \ $0.690$ & \ \ $0.690$\\
\hline
\end{tabular}\\
\end{minipage}
\caption{A tetrahedron that is completely well-centered and dihedral
  acute}
\label{fig:tetay2y3y}
\end{figure}

\begin{figure}[p]
\centering
\begin{minipage}[c]{160pt}
\includegraphics[width=150pt, trim=262pt 262pt 217pt 221pt, clip]
  {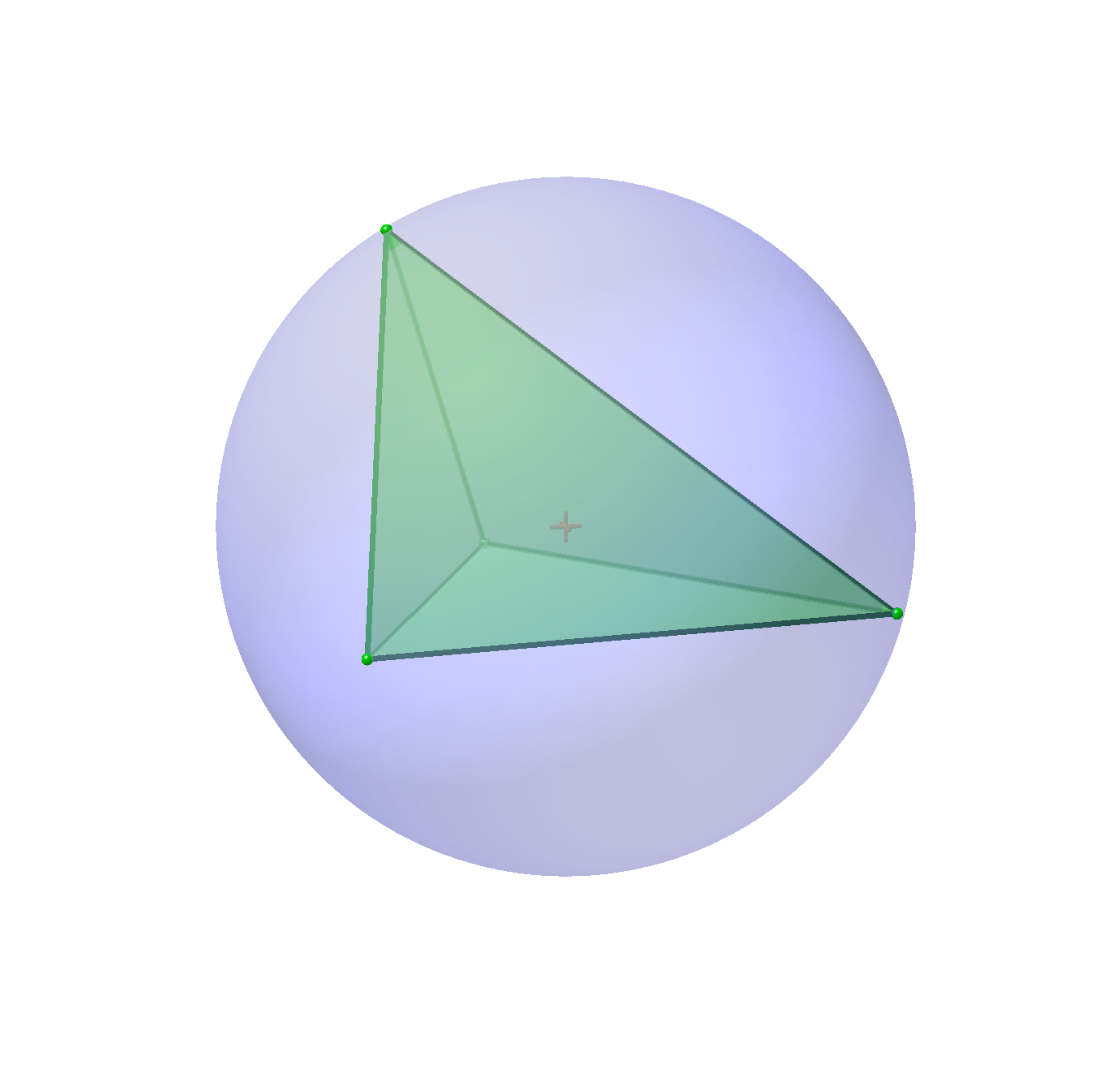}%
\end{minipage}%
\hspace{10pt}
\begin{minipage}[c]{200pt}
\centering
\begin{tabular}{|l|l|l|}
\hline
\multicolumn{3}{|c|}{Vertex Coordinates}\\
\hline
\multicolumn{1}{|c|}{$x$} & \multicolumn{1}{c|}{$y$}
  & \multicolumn{1}{c|}{$z$}\\
\hline
\ \ $\phantom{-}0$ & \ \ $\phantom{-}0.96$ & \ \ $-0.28$\\
\hline
\ \ $-0.744\phantom{0}$ & \ \ $-0.64\phantom{00}$
  & \ \ $-0.192\phantom{0}$\\
\hline
\ \ $\phantom{-}0.856$ & \ \ $-0.48$ & \ \ $-0.192$\\
\hline
\ \ $-0.48$ & \ \ $\phantom{-}0.192$ & \ \ $\phantom{-}0.856$\\
\hline
\end{tabular}\\[10pt]

\begin{tabular}{|l|l|l|}
\hline
\multicolumn{3}{|c|}{Quality Statistics}\\
\hline
\multicolumn{1}{|c|}{Quantity} & \multicolumn{1}{c|}{Min}
  & \multicolumn{1}{c|}{Max}\\
\hline
\ $h/R$ & \ \ $0.224$ & \ \ $0.427$\\
\hline
\ Face Angle & \ \ $46.26$\textdegree\ \  & \ \ $77.62$\textdegree\ \ \\
\hline
\ Dihedral Angle\ \ & \ \ $52.71$\textdegree & \ \ $94.15$\textdegree\\
\hline
\ $R/\ell$ & \ \ $0.733$ & \ \ $0.733$\\
\hline
\end{tabular}\\
\end{minipage}
\caption{A completely well-centered tetrahedron that is not dihedral acute}
\label{fig:tetan2y3y}
\end{figure}

\begin{figure}[p]
\centering
\begin{minipage}[c]{160pt}
\includegraphics[width=150pt, trim=197pt 196pt 150pt 155pt, clip]
  {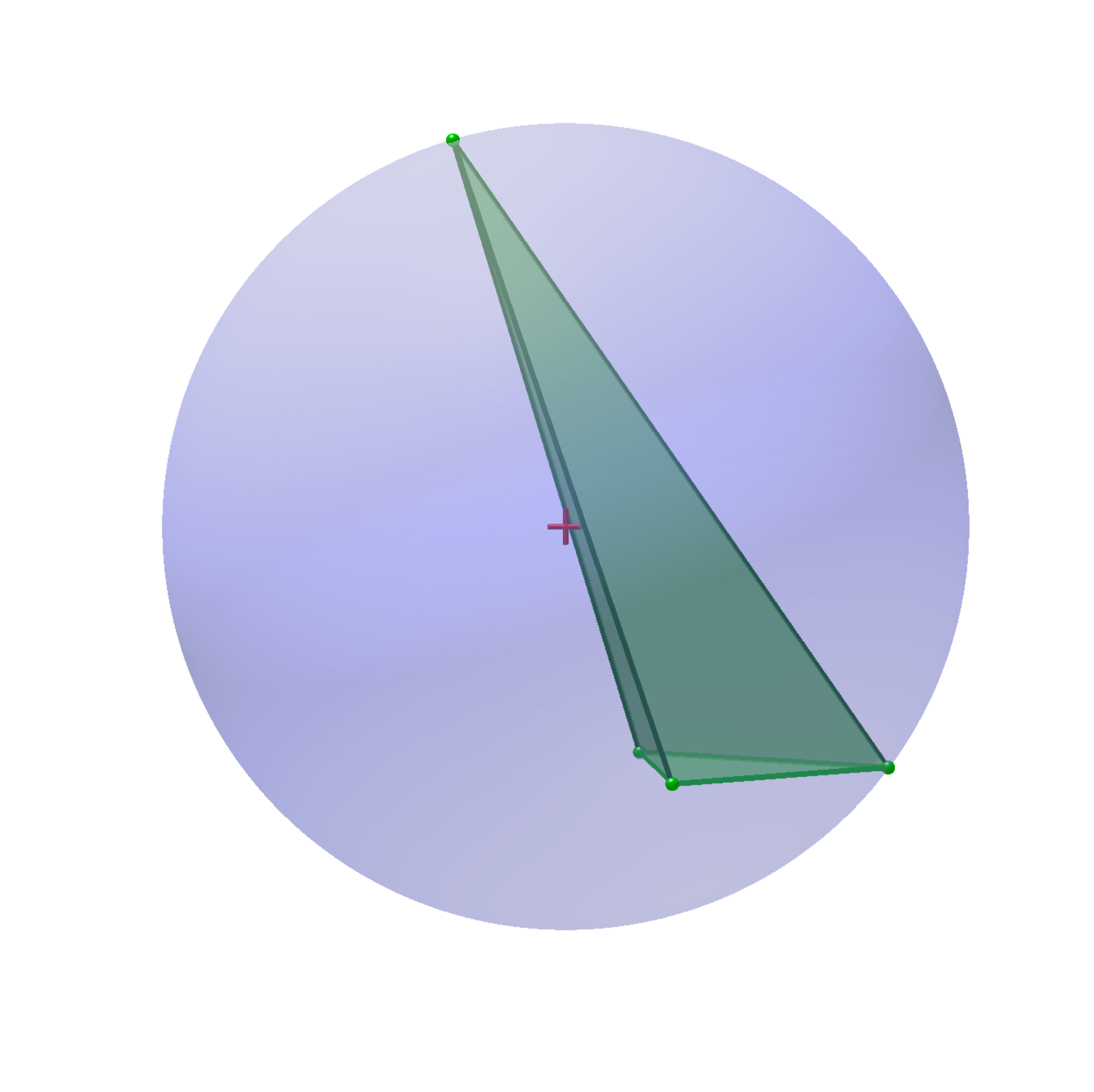}%
\end{minipage}%
\hspace{10pt}
\begin{minipage}[c]{200pt}
\centering
\begin{tabular}{|l|l|l|}
\hline
\multicolumn{3}{|c|}{Vertex Coordinates}\\
\hline
\multicolumn{1}{|c|}{$x$} & \multicolumn{1}{c|}{$y$}
  & \multicolumn{1}{c|}{$z$}\\
\hline
\ \ $\phantom{-}0.224\phantom{0}$ & \ \ $-0.768\phantom{0}$
  & \ \ $-0.6\phantom{000}$\\
\hline
\ \ $\phantom{-}0.8$ & \ \ $\phantom{-}0$ & \ \ $-0.6$\\
\hline
\ \ $\phantom{-}0.224$ & \ \ $\phantom{-}0.768$ & \ \ $-0.6$\\
\hline
\ \ $-0.28$ & \ \ $\phantom{-}0$ & \ \ $\phantom{-}0.96$\\
\hline
\end{tabular}\\[10pt]

\begin{tabular}{|l|l|l|}
\hline
\multicolumn{3}{|c|}{Quality Statistics}\\
\hline
\multicolumn{1}{|c|}{Quantity} & \multicolumn{1}{c|}{Min}
  & \multicolumn{1}{c|}{Max}\\
\hline
\ $h/R$ & \ $-0.029$ & \ \ $\phantom{0}0.600$\\
\hline
\ Face Angle & \ $\phantom{-}29.89$\textdegree\ \  & \ \ $106.26$\textdegree\ \ \\
\hline
\ Dihedral Angle\ \ & \ $\phantom{-}35.42$\textdegree & \ \ $116.68$\textdegree\\
\hline
\ $R/\ell$ & \ $\phantom{-}1.042$ & \ \ $\phantom{0}1.042$\\
\hline
\end{tabular}\\
\end{minipage}
\caption{A tetrahedron that is not dihedral acute,
  2-well-centered, or 3-well-centered}
\label{fig:tetan2n3n}
\end{figure}

\begin{figure}[p]
\centering
\begin{minipage}[c]{160pt}
\includegraphics[width=150pt, trim=321pt 321pt 274pt 276pt, clip]
  {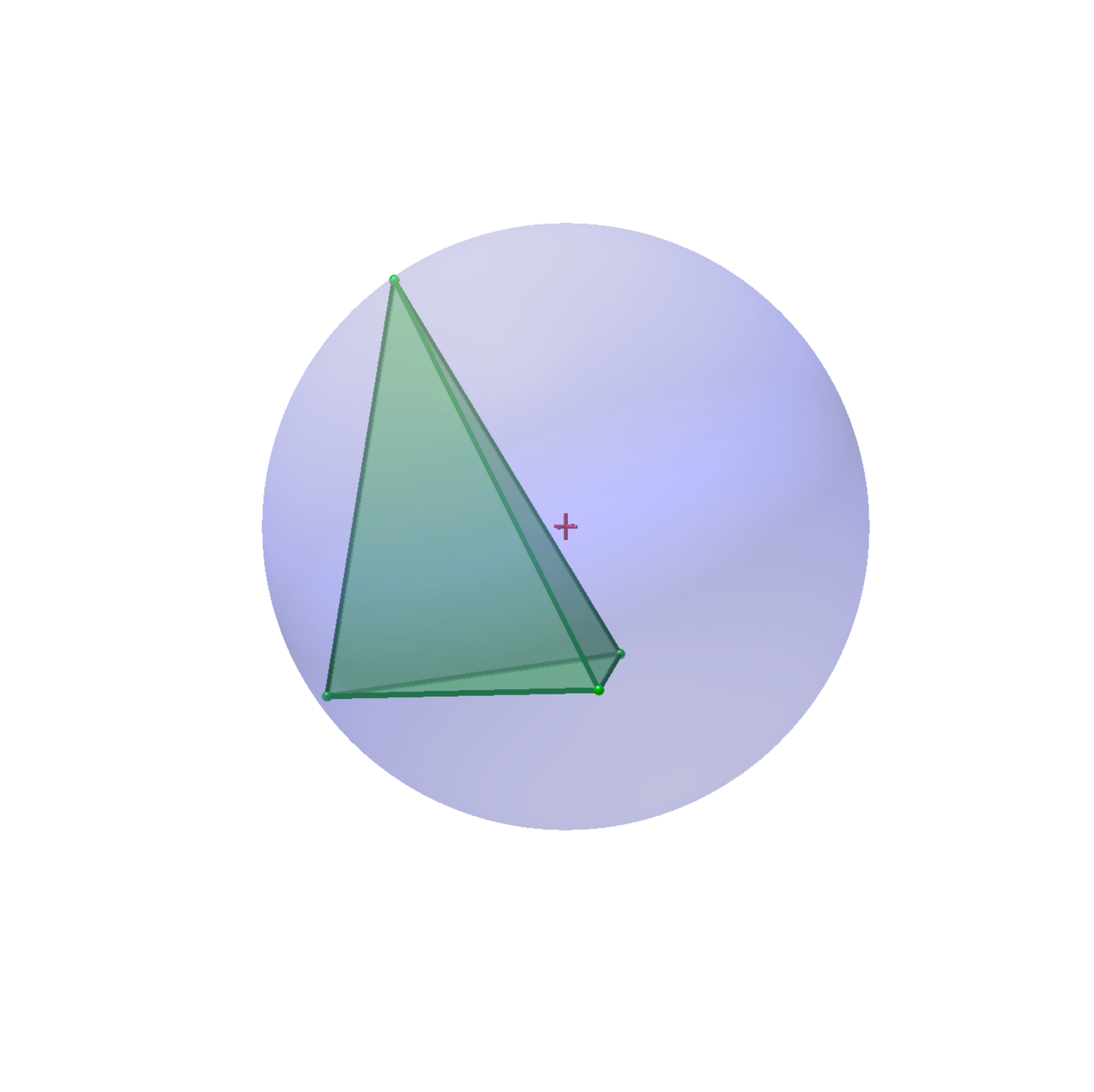}%
\end{minipage}%
\hspace{10pt}
\begin{minipage}[c]{200pt}
\centering
\begin{tabular}{|l|l|l|}
\hline
\multicolumn{3}{|c|}{Vertex Coordinates}\\
\hline
\multicolumn{1}{|c|}{$x$} & \multicolumn{1}{c|}{$y$}
  & \multicolumn{1}{c|}{$z$}\\
\hline
\ \ $\phantom{-}0.36\phantom{00}$ & \ \ $-0.8\phantom{000}$
  & \ \ $-0.48\phantom{00}$\\
\hline
\ \ $\phantom{-}0.768$ & \ \ $\phantom{-}0.28$ & \ \ $-0.576$\\
\hline
\ \ $-0.6$ & \ \ $\phantom{-}0.64$ & \ \ $-0.48$\\
\hline
\ \ $\phantom{-}0.576$ & \ \ $\phantom{-}0.168$ & \ \ $\phantom{-}0.8$\\
\hline
\end{tabular}\\[10pt]

\begin{tabular}{|l|l|l|}
\hline
\multicolumn{3}{|c|}{Quality Statistics}\\
\hline
\multicolumn{1}{|c|}{Quantity} & \multicolumn{1}{c|}{Min}
  & \multicolumn{1}{c|}{Max}\\
\hline
\ $h/R$ & \ $-0.109$ & \ \ $0.562$\\
\hline
\ Face Angle & \ $\phantom{-}41.71$\textdegree\ \  & \ \ $83.76$\textdegree\ \ \\
\hline
\ Dihedral Angle\ \ & \ $\phantom{-}53.33$\textdegree & \ \ $85.72$\textdegree\\
\hline
\ $R/\ell$ & \ $\phantom{-}0.863$ & \ \ $0.863$\\
\hline
\end{tabular}\\
\end{minipage}
\caption{A tetrahedron that is dihedral acute and 2-well-centered, but not
  3-well-centered}
\label{fig:tetay2y3n}
\end{figure}

\begin{figure}[p]
\centering
\begin{minipage}[c]{160pt}
\includegraphics[width=150pt, trim=287pt 287pt 240pt 241pt, clip]
  {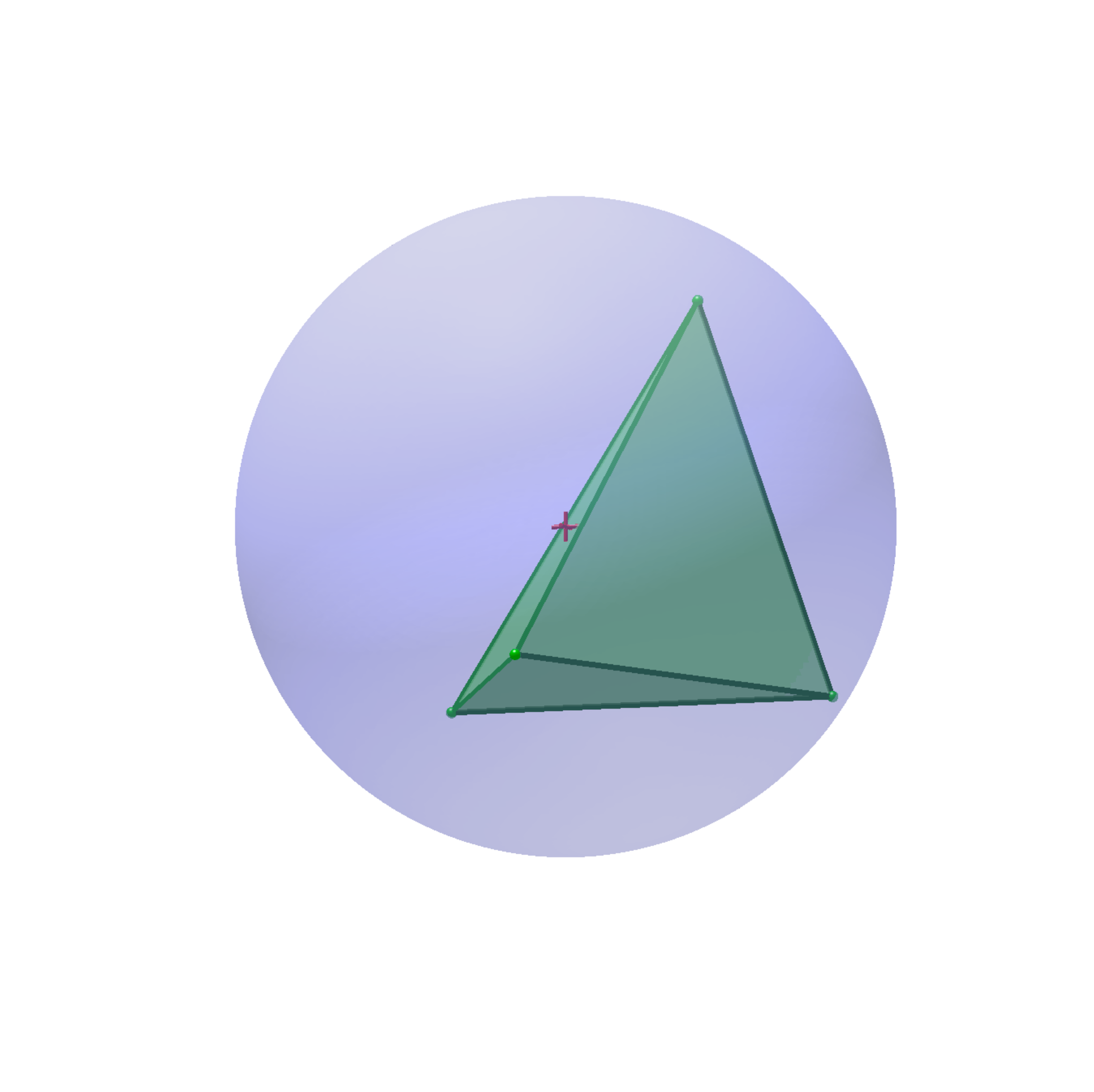}%
\end{minipage}%
\hspace{10pt}
\begin{minipage}[c]{200pt}
\centering
\begin{tabular}{|l|l|l|}
\hline
\multicolumn{3}{|c|}{Vertex Coordinates}\\
\hline
\multicolumn{1}{|c|}{$x$} & \multicolumn{1}{c|}{$y$}
  & \multicolumn{1}{c|}{$z$}\\
\hline
\ \ $-0.152\phantom{0}$ & \ \ $\phantom{-}0.864\phantom{0}$
  & \ \ $-0.48\phantom{00}$\\
\hline
\ \ $-0.64$ & \ \ $-0.6$ & \ \ $-0.48$\\
\hline
\ \ $\phantom{-}0.6$ & \ \ $-0.64$ & \ \ $-0.48$\\
\hline
\ \ $-0.192$ & \ \ $-0.64$ & \ \ $\phantom{-}0.744$\\
\hline
\end{tabular}\\[10pt]

\begin{tabular}{|l|l|l|}
\hline
\multicolumn{3}{|c|}{Quality Statistics}\\
\hline
\multicolumn{1}{|c|}{Quantity} & \multicolumn{1}{c|}{Min}
  & \multicolumn{1}{c|}{Max}\\
\hline
\ $h/R$ & \ $-0.024$ & \ \ $0.630$\\
\hline
\ Face Angle & \ $\phantom{-}42.08$\textdegree\ \  & \ \ $85.44$\textdegree\ \ \\
\hline
\ Dihedral Angle\ \ & \ $\phantom{-}59.94$\textdegree & \ \ $91.20$\textdegree\\
\hline
\ $R/\ell$ & \ $\phantom{-}0.806$ & \ \ $0.806$\\
\hline
\end{tabular}\\
\end{minipage}
\caption{A tetrahedron that is not dihedral acute or 3-well-centered,
  but is 2-well-centered}
\label{fig:tetan2y3n}
\end{figure}

\begin{figure}[p]
\centering
\begin{minipage}[c]{160pt}
\includegraphics[width=150pt, trim=312pt 312pt 264pt 265pt, clip]
  {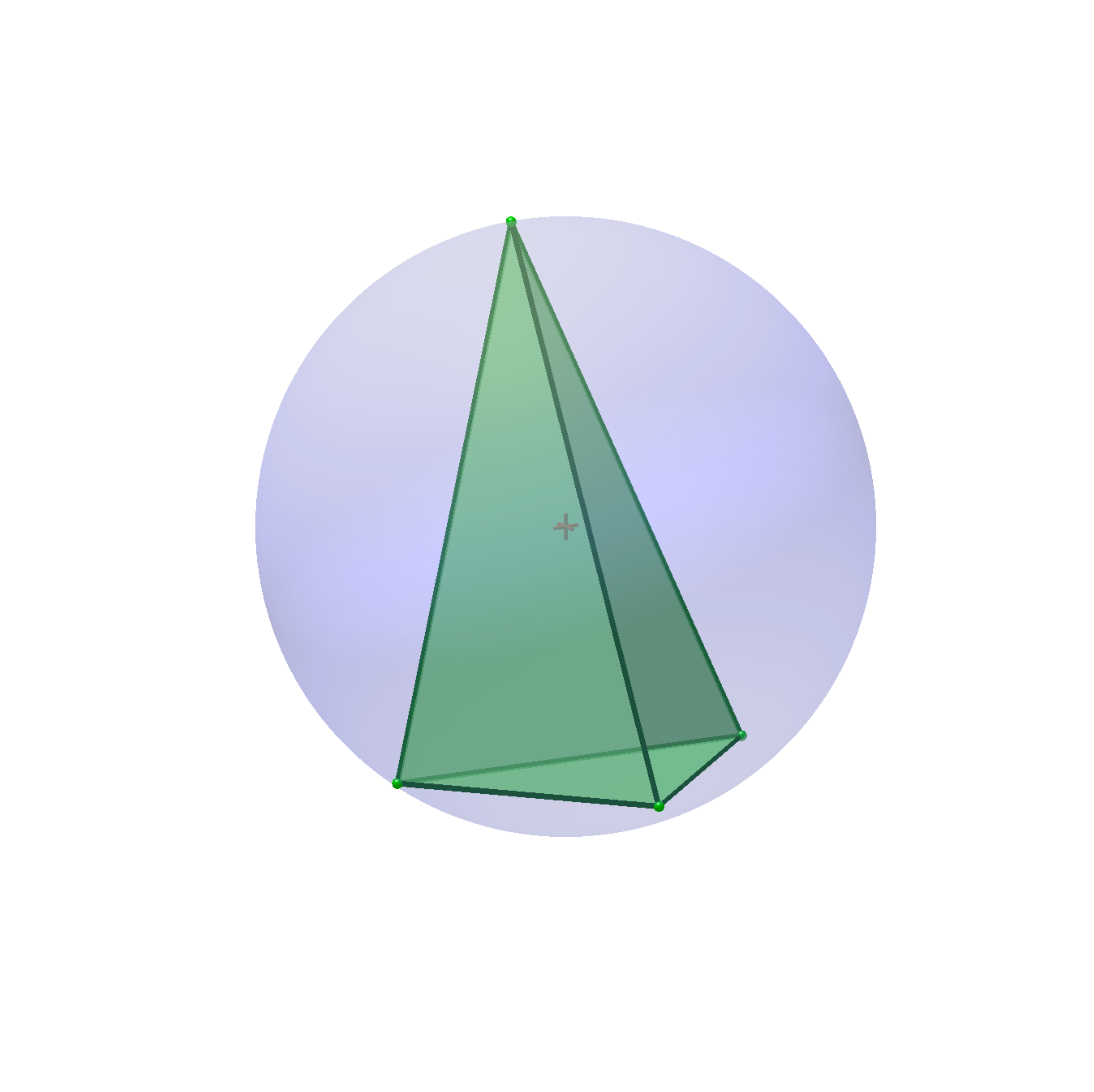}%
\end{minipage}%
\hspace{10pt}
\begin{minipage}[c]{200pt}
\centering
\begin{tabular}{|l|l|l|}
\hline
\multicolumn{3}{|c|}{Vertex Coordinates}\\
\hline
\multicolumn{1}{|c|}{$x$} & \multicolumn{1}{c|}{$y$}
  & \multicolumn{1}{c|}{$z$}\\
\hline
\ \ $\phantom{-}0$ & \ \ $-0.6$ & \ \ $-0.8$\\
\hline
\ \ $\phantom{-}0.64\phantom{00}$ & \ \ $-0.024\phantom{0}$
  & \ \ $-0.768\phantom{0}$\\
\hline
\ \ $-0.64$ & \ \ $-0.024$ & \ \ $-0.768$\\
\hline
\ \ $\phantom{-}0$ & \ \ $\phantom{-}0.352$ & \ \ $\phantom{-}0.936$\\
\hline
\end{tabular}\\[10pt]

\begin{tabular}{|l|l|l|}
\hline
\multicolumn{3}{|c|}{Quality Statistics}\\
\hline
\multicolumn{1}{|c|}{Quantity} & \multicolumn{1}{c|}{Min}
  & \multicolumn{1}{c|}{Max}\\
\hline
\ $h/R$ & \ \ $0.112$ & \ \ $\phantom{0}0.765$\\
\hline
\ Face Angle & \ \ $25.69$\textdegree\ \  & \ \ $\phantom{0}95.94$\textdegree\ \ \\
\hline
\ Dihedral Angle\ \ & \ \ $40.33$\textdegree & \ \ $105.62$\textdegree\\
\hline
\ $R/\ell$ & \ \ $1.161$ & \ \ $\phantom{0}1.161$\\
\hline
\end{tabular}\\
\end{minipage}
\caption{A tetrahedron that is not dihedral acute or 2-well-centered,
  but is 3-well-centered}
\label{fig:tetan2n3y}
\end{figure}

Figures~\ref{fig:tetay2y3y} through~\ref{fig:tetan2n3y} are pictures
of six different tetrahedra that illustrate the possible combinations
of the qualities $2$-well-centered, $3$-well-centered, and dihedral
acute.  Each picture shows a tetrahedron inside of its circumsphere.
The center of each circumsphere is marked by a small, unlabeled
axes indicator.  In each case, the circumcenter of the tetrahedron lies
at the origin, and the circumradius is $1$.  The coordinates given
are exact, and the quality statistics are rounded to the
nearest value of the precision shown.

The quality statistics displayed include the minimum and maximum face
angles and dihedral angles of the tetrahedron. These familiar quality
measures need no further explanation, and it is easy to determine
from them
whether a tetrahedron is $2$-well-centered (having all face angles acute) or
dihedral acute.  The $R/\ell$ statistic
shows the ratio of the circumradius $R$ to the shortest edge of the
tetrahedron, which has length $\ell$.  The range of 
$R/\ell$ is $[\sqrt{3/8},\infty]$, with $\sqrt{3/8} \approx 0.612$.
A single tetrahedron has a
particular $R/\ell$ ratio, so the minimum $R/\ell$ equals the
maximum $R/\ell$ in each of Figs.~\ref{fig:tetay2y3y}--\ref{fig:tetan2n3y}.
Later, however, we will show similar statistics for tetrahedral
meshes, and it is convenient to use the same format to summarize mesh
quality in both cases.
The $R/\ell$ ratio is a familiar measurement of the quality of a
tetrahedron, especially in the context of Delaunay refinement.

The quality statistic $h/R$ is less familiar.  The $R$ in this ratio
is the circumradius.  The $h$ stands for height.  For a given facet of
the tetrahedron, $h$ measures the signed height of the circumcenter of
the tetrahedron above the plane containing that facet.  The direction
above the facet means the direction towards the remaining vertex of
the tetrahedron, and $h$ is positive when the circumcenter lies above
the facet.  In \cite{VaHiGuRa2008} the quantity $h/R$ and its
relationship to well-centeredness in any dimension is discussed at
more length.  For our purposes it should suffice to note that the
range of $h/R$ for tetrahedra is $(-1, 1)$, that a
tetrahedron is $3$-well-centered if and only if the minimum $h/R$ is
positive, and that $h/R = 1/3$ relative to every
facet of the regular tetrahedron.

The regular tetrahedron is completely well-centered and dihedral
acute.  Our first example, shown in
Fig.~\ref{fig:tetay2y3y}, is another tetrahedron that
shares those properties with the regular tetrahedron.
Not every completely well-centered tetrahedron is dihedral acute,
and the second example, shown in Fig.~\ref{fig:tetan2y3y}, is a
tetrahedron that is completely well-centered but not dihedral acute.
Some sliver tetrahedra are completely well-centered and have
dihedral angles approaching $180$\textdegree.

As one might expect, there are tetrahedra that have none of these
nice properties.  The tetrahedron shown in
Fig.~\ref{fig:tetan2n3n} is an example of such a tetrahedron.  Most
polar caps also have none of these nice properties.  Some
tetrahedra that are neither dihedral acute nor $2$-well-centered
nor $3$-well-centered have much worse quality than the
example in Fig.~\ref{fig:tetan2n3n}.  This particular example
is near the boundary dividing $3$-well-centered tetrahedra from
tetrahedra that are not $3$-well-centered. 

The tetrahedra we have considered so far have been either completely
well-centered or neither 2-well-centered nor 3-well-centered.  Our
last three examples show that tetrahedra can be $2$-well-centered
without being $3$-well-centered and vice-versa.  The
tetrahedron shown in Fig.~\ref{fig:tetay2y3n} is $2$-well-centered
and dihedral acute, but not $3$-well-centered.
The example tetrahedron shown in Fig.~\ref{fig:tetan2y3n} is similar
but has been modified to no longer be dihedral acute.
Our final example, shown in Fig.~\ref{fig:tetan2n3y},
is a tetrahedron that is 3-well-centered,
but is neither 2-well-centered nor dihedral acute.

\begin{table}
\centering
\begin{tabular}{|c|c|c|c|}
\hline
\ Figure\ \ &\ 3-WC\ \ &\ 2-WC\ \ &\ Acute\ \ \\
\hline
\ref{fig:tetay2y3y} & Y & Y & Y\\
\hline
\ref{fig:tetan2y3y} & Y & Y & N\\
\hline
\ref{fig:tetan2n3n} & N & N & N\\
\hline
\ref{fig:tetay2y3n} & N & Y & Y\\
\hline
\ref{fig:tetan2y3n} & N & Y & N\\
\hline
\ref{fig:tetan2n3y} & Y & N & N\\
\hline
\end{tabular}
\caption{A tetrahedron may have any of six different possible
  combinations of the qualities $2$-well-centered, $3$-well-centered,
  and dihedral acute}
\label{table:examplesummary}
\end{table}

Table~\ref{table:examplesummary} summarizes the six examples presented
in this section, indicating whether each example is $2$-well-centered,
$3$-well-centered, and/or dihedral acute.  Of the eight possible
binary sequences, the two missing examples are the sequences $N, N, Y$
and $Y, N, Y$, in which a tetrahedron would be dihedral acute but not
$2$-well-centered.  These examples are missing because they do not
exist; every tetrahedron that is dihedral acute is also
2-well-centered.  Eppstein, Sullivan, and \"Ung\"or provide a proof of
this in Lemma 2 of \cite{EpSuUn2004}, which states, among other
things, that ``an acute tetrahedron has acute facets."

\section{Tiling Space, Slabs, and Infinite Rectangular Prisms \\
  with Completely Well-Centered Tetrahedra}

We have mentioned above that there are applications that could make
good use of well-centered triangulations and have considered examples of
single well-centered tetrahedra. In what follows, we give some examples
of well-centered triangulations of simple
domains in dimension $3$.

In \cite{EpSuUn2004}, Eppstein, Sullivan, and \"Ung\"or show that one
can tile space, $\RR^3$, and infinite slabs,
$\RR^2\times[0,a]$, with dihedral acute
tetrahedra.  They also briefly discuss how high-quality tilings of
space have been used to design meshing algorithms.  The acute
triangulations of space given in \cite{EpSuUn2004} all make use of copies
of at least two different tetrahedra, and the authors suggest it is
unlikely that there is a tiling of space with copies of a single acute
tetrahedron.  Their acute triangulation of the slab appears to use
copies of seven distinct tetrahedra.  The problem of triangulating an
infinite rectangular prism, $\RR \times [0,a] \times [0, b]$,
or a cube, $[0,1]^3$, with acute tetrahedra is still an open
problem as far as the authors know.

\begin{figure}
\centering
\begin{minipage}[c]{160pt}
\includegraphics[width=150pt, trim=228pt 228pt 182pt 183pt, clip]
  {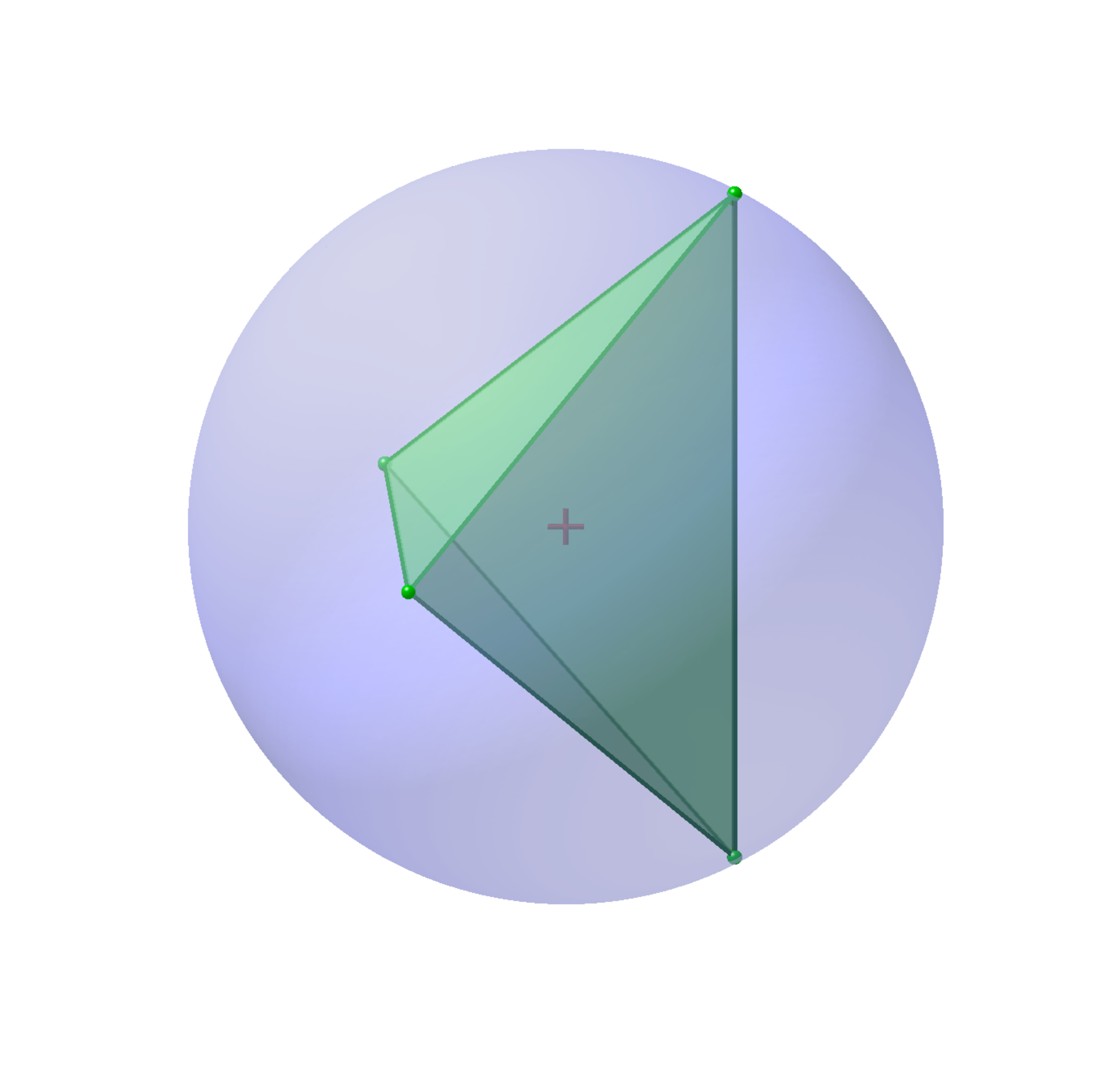}%
\end{minipage}%
\hspace{10pt}
\begin{minipage}[c]{200pt}
\centering
\begin{tabular}{|l|l|l|}
\hline
\multicolumn{3}{|c|}{Vertex Coordinates}\\
\hline
\multicolumn{1}{|c|}{$x$} & \multicolumn{1}{c|}{$y$}
  & \multicolumn{1}{c|}{$z$}\\
\hline
\ \ $-1\phantom{0}$ & \ \ $-2\phantom{0}$
  & \ \ $\phantom{-}0\phantom{0}$\\
\hline
\ \ $\phantom{-}1$ & \ \ $\phantom{-}0$ & \ \ $-2$\\
\hline
\ \ $-1$ & \ \ $\phantom{-}2$ & \ \ $\phantom{-}0$\\
\hline
\ \ $\phantom{-}1$ & \ \ $\phantom{-}0$ & \ \ $\phantom{-}2$\\
\hline
\end{tabular}\\[10pt]

\begin{tabular}{|l|l|l|}
\hline
\multicolumn{3}{|c|}{Quality Statistics}\\
\hline
\multicolumn{1}{|c|}{Quantity} & \multicolumn{1}{c|}{Min}
  & \multicolumn{1}{c|}{Max}\\
\hline
\ $h/R$ & \ $0.316$ & \ \ $0.316$\\
\hline
\ Face Angle & \ $54.74$\textdegree\ \  & \ \ $70.53$\textdegree\ \ \\
\hline
\ Dihedral Angle\ \ & \ $60.00$\textdegree & \ \ $90.00$\textdegree\\
\hline
\ $R/\ell$ & \ $0.645$ & \ \ $0.645$\\
\hline
\end{tabular}\\
\end{minipage}
\caption{The second Sommerville tetrahedron -- a completely
  well-centered tetrahedron that tiles space}
\label{fig:spacetile}
\end{figure}

In contrast to the complexity of tiling space with acute tetrahedra,
there are fairly simple completely well-centered triangulations of
space.  Barnes and Sloane proved that the optimal lattice for
quantizing uniformly distributed data in $\RR^3$ is the body-centered
cubic (BCC) lattice \cite{BaSl1983}.  Since this is related to
centroidal Voronoi tesselations (CVTs) \cite{DuFaGu1999}, and CVTs
have been used for high-quality meshing of $3$-dimensional domains
(see \cite{AlCoYvDe2005}), it is not surprising that a Delaunay
triangulation of the vertices of the BCC lattice gives rise to a high
quality triangulation of space.  The triangulation consists of
congruent copies of a single completely well-centered tetrahedron,
shown in Fig.~\ref{fig:spacetile}.  The tiling is one of four spatial
tilings discovered by Sommerville \cite{Sommerville1923,EpSuUn2004}.
The other three tilings Sommerville found are neither
$3$-well-centered nor $2$-well-centered, though one of them has
a maximum face angle of $90$\textdegree\ and is dihedral nonobtuse.
It is interesting to note that Fuchs algorithm for meshing
spatial domains based on high-quality spatial tilings had ``good
performance \ldots when he used the second Sommerville
construction,"~S\cite{EpSuUn2004,Fuchs1998}
which is the completely well-centered tetrahedron
shown in Fig.~\ref{fig:spacetile}.

Next we describe the space tiling that uses the Sommerville
tetrahedron described above.  The view of the tetrahedron shown in
Fig.~\ref{fig:spacetile} has an elevation angle between $10$ and $11$
degrees, so it is not difficult to identify the horizontal edge and
the vertical edge of the tetrahedron.  The horizontal edge connects a
pair of vertices from one of the cubic lattices, and the vertical edge
connects a pair of vertices from the other cubic lattice.  It is
natural to think of
this spatial tiling in terms of the interleaved cubic lattices, but we
can also look at the tiling from a different perspective. Consider
the tilted plane $P$ that contains the bottom face of the tetrahedron
shown in Fig.~\ref{fig:spacetile}.  Consider a single cube of one of
the cubic lattices. The plane $P$ is the one defined by two opposing
parallel edges of the cube.  At the center of the cube there is
a vertex from the other cube lattice;  this vertex also lies in $P$.
By making translated copies of the vertices in
$P$ we can obtain all of the vertices of the BCC lattice.  Each vertex
in $P$ has six adjacent vertices in $P$ and four adjacent vertices in
each of the plane above and below $P$.  There are two types of
tetrahedra in the tiling, but both types are copies of the same
tetrahedron in this case. The first type is the convex hull of three
vertices in a copy of $P$ and one vertex from the plane above or below
that copy of $P$.  The second type of tetrahedron is the convex hull
of an edge in a copy of $P$ and a corresponding edge from the plane
above or below that copy of $P$.


From this understanding of the structure of the BCC-based spatial
tiling, we can generalize to an entire family of triangulations of
space using copies of two different tetrahedra.  We consider first the
set of vertices $\{(i, 0, 0):i\in\ZZ\}$.
We will make translated copies of this line in
one direction to make a plane $P$.  To do this we choose a parameter
$a > 0$ and make infinitely many copies of each vertex translating by
the vector $(1/2, a, 0)$.  Lastly we choose a parameter $b > 0$ and
make translated copies of the plane $P$ using the translation vector
$(1/2, 0, b)$.  Thus our set of vertices is
\[
\{(i + j/2 + k/2, aj, bk): i,j,k\in\ZZ\},
\]
i.e., the lattice
$\Lambda = \{\sum_{i=1}^{3} c_{i}u_{i}:c_{i} \in \ZZ\}$
with basis vectors
\[
u_{1} = (1, 0, 0), u_{2} = (1/2, a, 0),\ \textrm{and}\ u_{3} = (1/2, 0, b).
\] 

\begin{figure}
\centering
\includegraphics[width=306pt, trim=1265pt 405pt 775pt 1035pt, clip]%
  {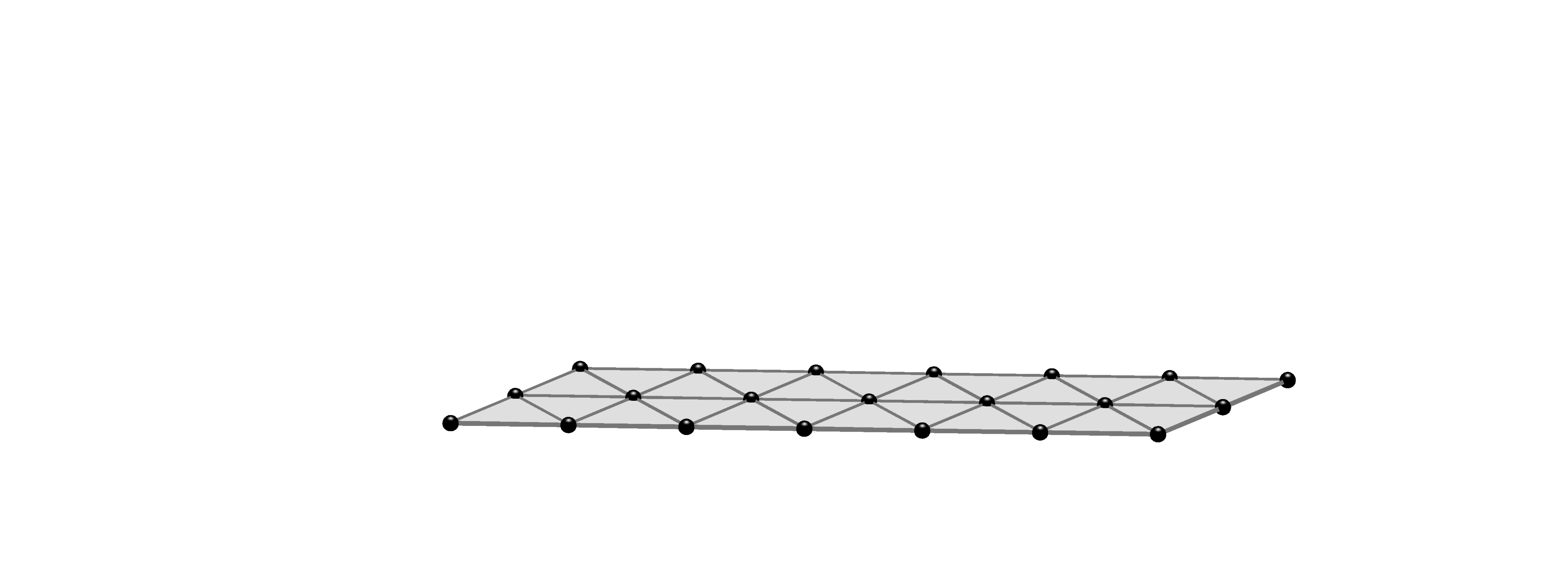}%
\caption{Starting with a set of vertices equally spaced along a line,
  we make translated copies of the line in a plane
  and triangulate the plane with each vertex having six neighbors}
\label{fig:spacetilefamplane}
\end{figure}

\begin{figure}
\centering
\includegraphics[width=306pt, trim=1265pt 405pt 435pt 385pt, clip]%
  {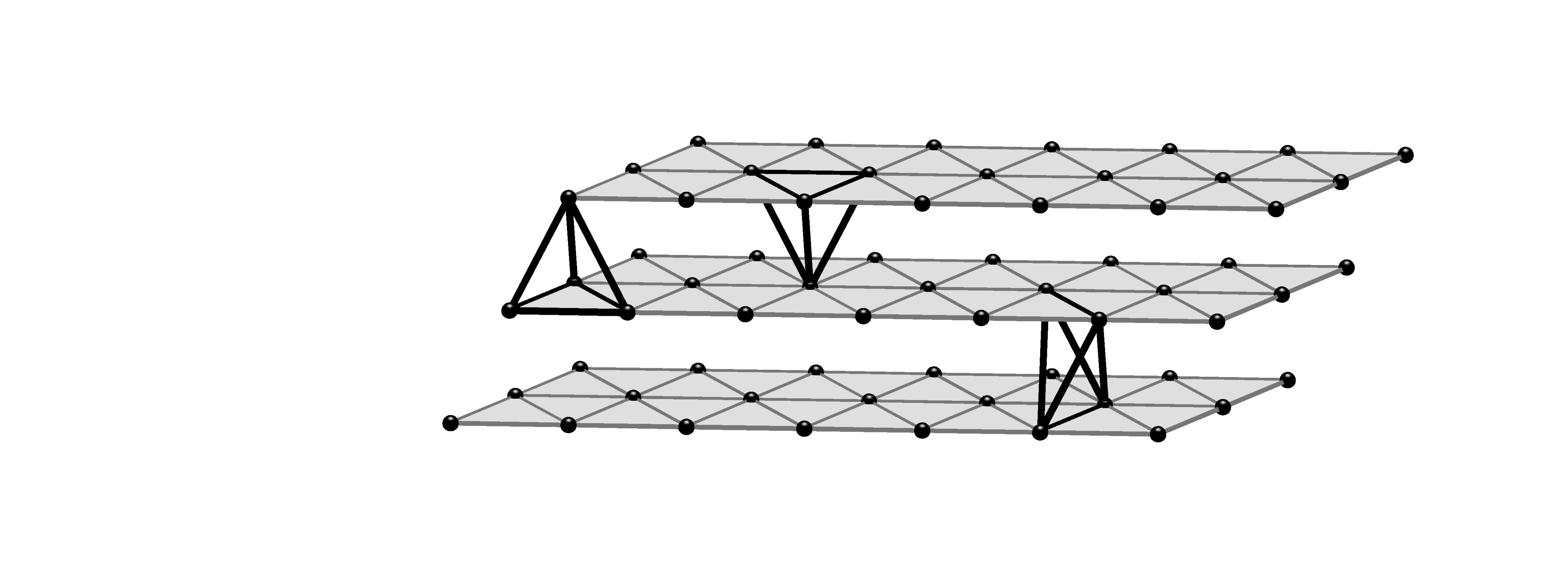}%
\caption{We make translated copies of the plane into space and form
  tetrahedra by connecting vertices of adjacent planes as shown}
\label{fig:spacetilefamtets}
\end{figure}

To turn this into a triangulation of space, we start by triangulating
each copy of the plane $P$ as shown in Fig.~\ref{fig:spacetilefamplane}.
Each vertex $v$ in a copy of $P$ is
connected to the six vertices which lie at positions $v \pm (1, 0,
0)$, $v \pm (1/2, a, 0)$, and $v \pm (1/2, -a, 0)$.
This yields the standard tiling of the plane with equilateral
triangles if $a = \sqrt{3}/2$.  Now for each triangle in a copy of $P$
there is exactly one edge -- the edge in the
direction $(1, 0, 0)$ -- for which a vertex lies
directly above and directly below the midpoint of that
edge. The first type of tetrahedron, then, is the convex hull of a
triangle $T$ in $P$ and the vertex directly above or below the
midpoint of one of the edges of $T$.  If we add all possible
tetrahedra of the first type, the gaps that remain are all tetrahedra
of the second type.  This second type of tetrahedron is the convex
hull of an edge in $P$ in the direction $\pm(1/2, a, 0)$ and the edge
in the direction $\pm(1/2, -a, 0)$ whose midpoint lies directly above or below
the midpoint of the given edge.
Figure~\ref{fig:spacetilefamtets} shows three copies of
the plane and highlights three of the tetrahedra that appear
in the spatial tiling.  The two tetrahedra on the left side of
Fig.~\ref{fig:spacetilefamtets} are both copies of the first type
of tetrahedron.  The other tetrahedron is of the second type.
For the BCC-based spatial triangulation, in which
both types of tetrahedra are the same, the parameters are $a = b =
\sqrt{2}/2$.

This family of triangulations of space is interesting for at
least two reasons.  First, the triangulation is completely well-centered
if and only if both $a > 1/2$ and $b > 1/2.$
Second, the family provides an elegant solution to the problems of
tiling an infinite slab in $\RR^3$ and tiling infinite rectangular
prisms in $\RR^3$.  To tile the slab, one uses a finite number of
translates of plane $P$.  We see that any slab can be triangulated
using copies of a single tetrahedron; for the parameters $a = b =
\sqrt{2}/2,$ the two types of tetrahedra are the same, and we can
scale the result as needed to get a slab of the desired thickness.
Triangulating rectangular prisms is also easy; it suffices to use a
finite number of translates both of the initial line and
of the resulting infinite strip.  Again, this can
be done using copies of a single tetrahedron, provided that the ratio
of side lengths of the rectangle is a rational number.  If the ratio
of side lengths is $p/q$, one can use parameters $a = b = \sqrt{2}/2$
and take $p$ copies of the initial line with $q$ copies of the
infinite strip.  The result has the correct ratio of side lengths and
can be scaled to get the desired rectangle.

\section{Meshing the Cube With Well-Centered Tetrahedra}

Meshing the cube with well-centered tetrahedra is significantly more
difficult than meshing an infinite square prism, but it can be done.
In this section we discuss several well-centered meshes of the cube
that we constructed.  In each case, the mesh was built by first
designing the mesh connectivity, then moving the internal vertices
using the optimization algorithm described in \cite{VaHiGuRa2008}.

The first completely well-centered mesh of the cube that we discovered
has 224 tetrahedra.  We do not discuss the construction of this mesh
in detail here, but it is worth mentioning the mesh because it has
higher quality than the other well-centered meshes of the cube we will
discuss.  The quality statistics for the mesh are shown in the columns
to the left in Table~\ref{table:cubestats}.  The faces of this
triangulation of the cube match up with each other, so it is
relatively easy to make a well-centered mesh of any figure that can be
tiled with unit cubes.  One can use a copy of this well-centered mesh
of the cube in each cube tile, using rotations and reflections as
needed to make all the faces match.

Cassidy and Lord showed that the smallest acute triangulation of the
square consists of eight triangles \cite{CaLo1980}.  Knowing that a
completely well-centered mesh of the cube exists, it is natural to ask
the analogous question for the cube.  What is the smallest
well-centered mesh of the cube?  One can ask this question for the
three different types of well-centered tetrahedral meshes -- $2$-well-centered,
$3$-well-centered, and completely well-centered.  It is also conceivable
that the well-centered mesh of the cube with the fewest tetrahedra is
different from the mesh of the cube with the fewest vertices or edges,
but we restrict our attention to the smallest well-centered mesh in
the sense of fewest tetrahedra.

\begin{figure}
\centering
\includegraphics[width=320pt, trim=30pt 10pt 180pt 10pt, clip]
  {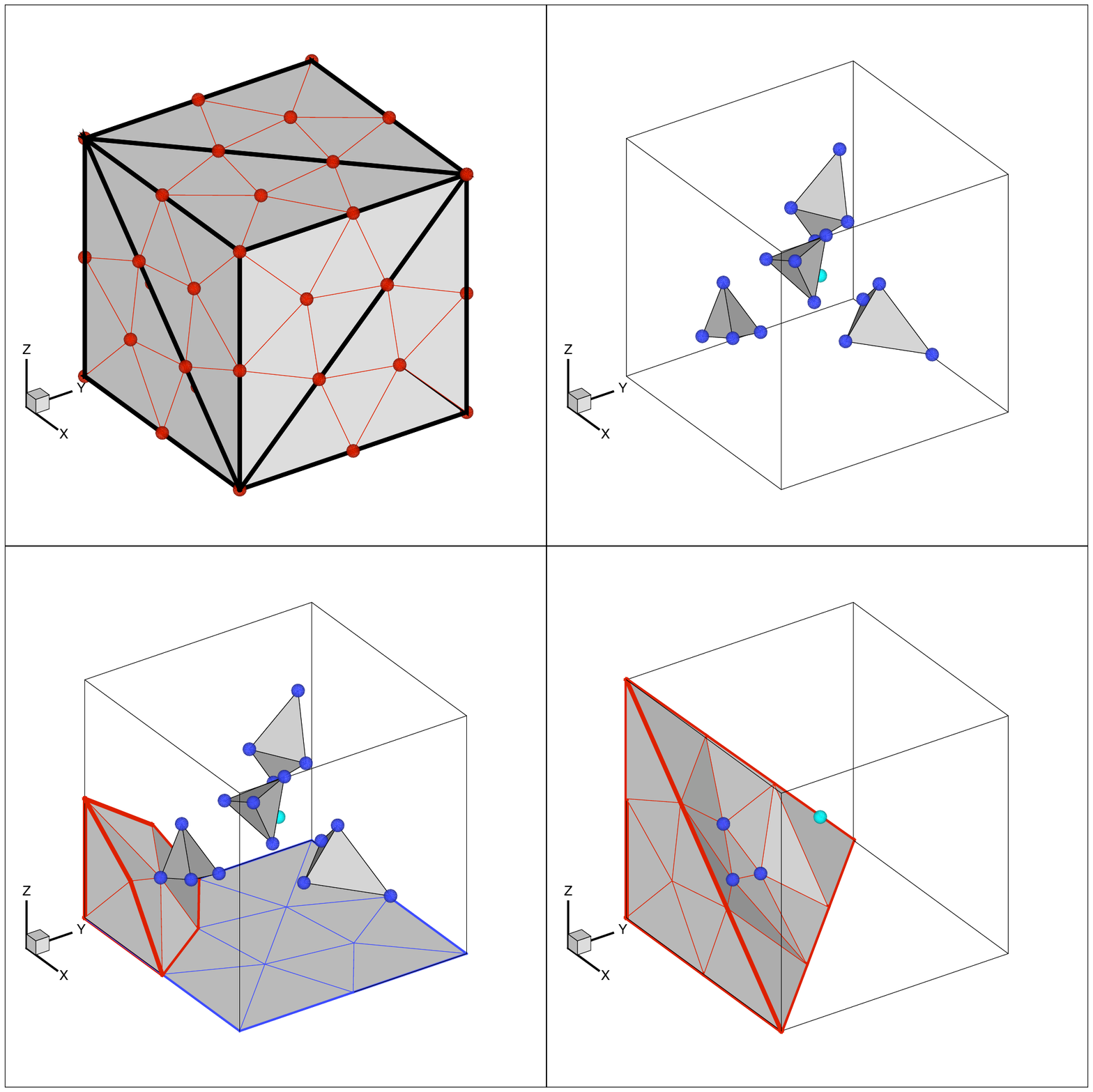}
\caption{A completely well-centered mesh of the cube with 194 tetrahedra}
\label{fig:cube_194_hsbcwc}
\end{figure}

The answer to the question is not known for any of the three types
of well-centered triangulations, but we can give
some upper and lower bounds.  The best known upper bound for
$2$-well-centered and completely well-centered triangulations of
the cube is a completely well-centered mesh of the cube with 194
tetrahedra.  Figure~\ref{fig:cube_194_hsbcwc} shows a picture of
this mesh.  The quality statistics for the mesh are recorded in
the middle columns of Table~\ref{table:cubestats}.

It is possible to improve the quality slightly by
optimization of the location of the surface vertices, but
for the version of the mesh shown in
Fig.~\ref{fig:cube_194_hsbcwc}, the surface triangulation
has some desirable symmetries, with every surface vertex at a
cube corner, at the midpoint of an edge of the cube,
or on a diagonal of a face of the cube.
All of the faces match each other up
to rotation and reflection, so this triangulation of the cube
can also be used to create a well-centered mesh of any figure
that can be tiled with unit cubes.  Although the surface
triangulation of this mesh is combinatorially the same as the
triangulation with 224 tetrahedra, the vertices on the surfaces
are not in the same location, so it is not trivial to mix these
two triangulations in meshing a cube-tiled shape.
It is worth noting that the vertices of the surface triangulation
that lie on the cube diagonals do not have coordinates of
$1/3$ or $2/3$.  Instead the coordinates are $0.35$ or $0.65$
for the vertices adjacent to an edge through the center of the
cube face, and the coordinates are $0.295$ or $0.705$ for vertices
not adjacent to such an edge.

\begin{table}
\centering
\begin{tabular}{|l|l|l|l|l|l|l|}
\hline
\multicolumn{7}{|c|}{Cube Mesh Quality Statistics}\\
\hline
\multicolumn{1}{|c|}{\multirow{2}{*}{Quantity}}
& \multicolumn{2}{c|}{224 Tets} &
  \multicolumn{2}{c|}{194 Tets} & \multicolumn{2}{c|}{146 Tets}\\
\cline{2-7}
& \multicolumn{1}{c|}{Min}
  & \multicolumn{1}{c|}{Max} & \multicolumn{1}{c|}{Min}
  & \multicolumn{1}{c|}{Max} & \multicolumn{1}{c|}{Min}
  & \multicolumn{1}{c|}{Max}\\
\hline
\ $h/R$ & \ $0.041$ & \ \ $\phantom{0}0.850$
  & \ $0.005$ & \ \ $\phantom{0}0.790$
  & \ $0.016$ & \ \ $\phantom{0}0.854$\\
\hline
\ Face Angle & \ $21.01$\textdegree\ \ &
  \ \ $\phantom{0}87.49$\textdegree\ \ 
  & \ $26.93$\textdegree\ \ &
    \ \ $\phantom{0}89.61$\textdegree\ \ 
  & \ $17.09$\textdegree\ \ &
  \ \ $112.60$\textdegree\ \ \\
\hline
\ Dihedral Angle\ \ & \ $24.91$\textdegree & \ \ $105.61$\textdegree
  & \ $28.26$\textdegree & \ \ $126.64$\textdegree
  & \ $10.73$\textdegree & \ \ $163.17$\textdegree\\
\hline
\ $R/\ell$ & \ $0.618$ & \ \ $\phantom{0}1.569$
  & \ $0.612$ & \ \ $\phantom{0}1.134$
  & \ $0.711$ & \ \ $\phantom{0}1.835$\\
\hline
\end{tabular}\\
\caption{The quality of our meshes of the cube decreases with the
  number of tetrahedra in the mesh of the cube}
\label{table:cubestats}
\end{table}
%
%

Figure~\ref{fig:cube_194_hsbcwc} is designed to make it possible
to discern the structure of this triangulation of the cube.  In
the bottom right we see a triangulation of a region that fits
into the corner of the cube and extends along the cube surface
to diagonals of three of the faces of the cube.  The triangulation basically
has two layers of tetrahedra.  The first layer is shown at bottom
left.  It consists of six tetrahedra that are incident to the corner
of the cube.  The two-layer triangulation is imprecisely replicated
in four different corners of the cube.  The thicker diagonal lines
in the picture at top left help show which cube corners contain this type
of triangulation.  To complete the triangulation of the cube,
a vertex at the center of the cube is added to the triangulations
of the cube corners.  The picture at top right of
Fig.~\ref{fig:cube_194_hsbcwc} shows all the interior vertices of
the cube, combining the vertex at the center of the cube
with one tetrahedron from each triangulation of a cube corner.
The Delaunay triangulation defines the connectivity table, since
any $3$-well-centered triangulation is Delaunay.

This mesh establishes upper bounds for the $2$-well-centered
and completely well-centered cases.  There is an even smaller
triangulation of the cube that is known to be $3$-well-centered.
We do not describe it here except to say that the mesh consists
of 146 tetrahedra and has a
surface triangulation with fewer triangles than
the two previously mentioned triangulations of the cube.  The
quality of that $3$-well-centered mesh of the cube is
shown in the rightmost columns of Table~\ref{table:cubestats}.

The upper bounds for this problem are simple to present, since they
consist of constructive examples.  The analytical lower bounds are
rather more complicated to explain, and we discuss them here only
briefly.  One can show that in a $3$-well-centered triangulation of
the cube, no face of the cube is triangulated with two right triangles
meeting along the hypotenuse \cite{VaHiGuZh2008}.
Thus each face of the cube must contain
at least five vertices and at least three tetrahedral facets.  Since
there are six cube faces, this leads to a count of $18$ tetrahedra,
one adjacent to each facet.  Some of these tetrahedra may have been
counted twice, though, since a single tetrahedron may have a facet in
each of two different cube faces.  It can be shown that no tetrahedron
having a facet in each of three different cube faces is a
$3$-well-centered tetrahedron \cite{VaHiGuZh2008}.  Thus
no tetrahedron is triple-counted.  A lower bound
of $9$ tetrahedra follows.  It is possible to prove that any
triangulation of the cube with
all vertices lying on the edges of the cube is not $3$-well-centered,
but it is not immediately clear whether this can be used to improve
the lower bound.

The lower bound for $2$-well-centered and completely well-centered
triangulations of the cube is slightly better.  In this case, each
face of the cube must be triangulated with an acute triangulation, so
each cube face contains at least $8$ tetrahedral facets.  This gives a
count of $48$ tetrahedra, and once again we cannot have counted any
tetrahedron three times \cite{VaHiGuZh2008}.
(For the $2$-well-centered case this is not
exactly the same reason as the $3$-well-centered case.)  A lower bound
of $24$ tetrahedra follows.  This lower bound can be improved a little
by paying attention to which triangular facets can and cannot lead to
double-counted tetrahedra, and a lower bound of $30$ tetrahedra can be
obtained.

A careful analysis along these lines might improve the lower bound
even more, since there is no way to conformally triangulate all the
surfaces of the cube with an $8$-triangle acute triangulation of each
face.  In any case, the authors suspect that the actual answer to
these questions is close to $100$ tetrahedra, if not greater, so the
lower bounds mentioned here should be considered preliminary.

\section{Some Subdivisions of Tetrahedra
  into Well-Centered Tetrahedra}

Having seen that the cube can be subdivided into well-centered
tetrahedra, one might also ask whether a tetrahedron can be
subdivided into well-centered tetrahedra.  Subdivisions of tetrahedra
into well-centered tetrahedra could be used to refine an existing
mesh.  Subdividing a tetrahedron that tiles space into well-centered
tetrahedra would provide new well-centered meshes of space.  In general,
well-centered subdivisions of tetrahedra might be used to design
high-quality meshing algorithms.

One might suppose that subdividing the regular tetrahedron into
smaller well-centered tetrahedra is relatively simple, but the
problem is not so easy as it might seem.  In two dimensions,
the Loop subdivision, which refines a triangle by connecting
the midpoints of each edge of the triangle, produces
four smaller triangles.  Each of these triangles is
similar to the original triangle, so the Loop subdivision
of an acute triangle is an acute triangulation of the
triangle.  In three dimensions, however, there is no obvious
analog of the Loop subdivision.

Connecting the midpoints of the edges of a tetrahedron cuts
out four corner tetrahedra that are similar to the original
tetrahedron.  The shape that remains in the center after
removing these four tetrahedra is an octahedron.  In the
case of the regular tetrahedron, it is
a regular octahedron, and it can be subdivided
into four tetrahedra by adding an edge between opposite
vertices of the octahedron.  The result is not well-centered;
the center of the octahedron is the circumcenter of all four
tetrahedra, so the tetrahedra are not $3$-well-centered.  In
addition, the facets incident to the new edge are right
triangles, so the tetrahedra are not $2$-well-centered.

\begin{figure}
\centering
\begin{minipage}[c]{160pt}
\includegraphics[width = 153pt, trim = 322pt 108pt 146pt 115pt, clip]%
  {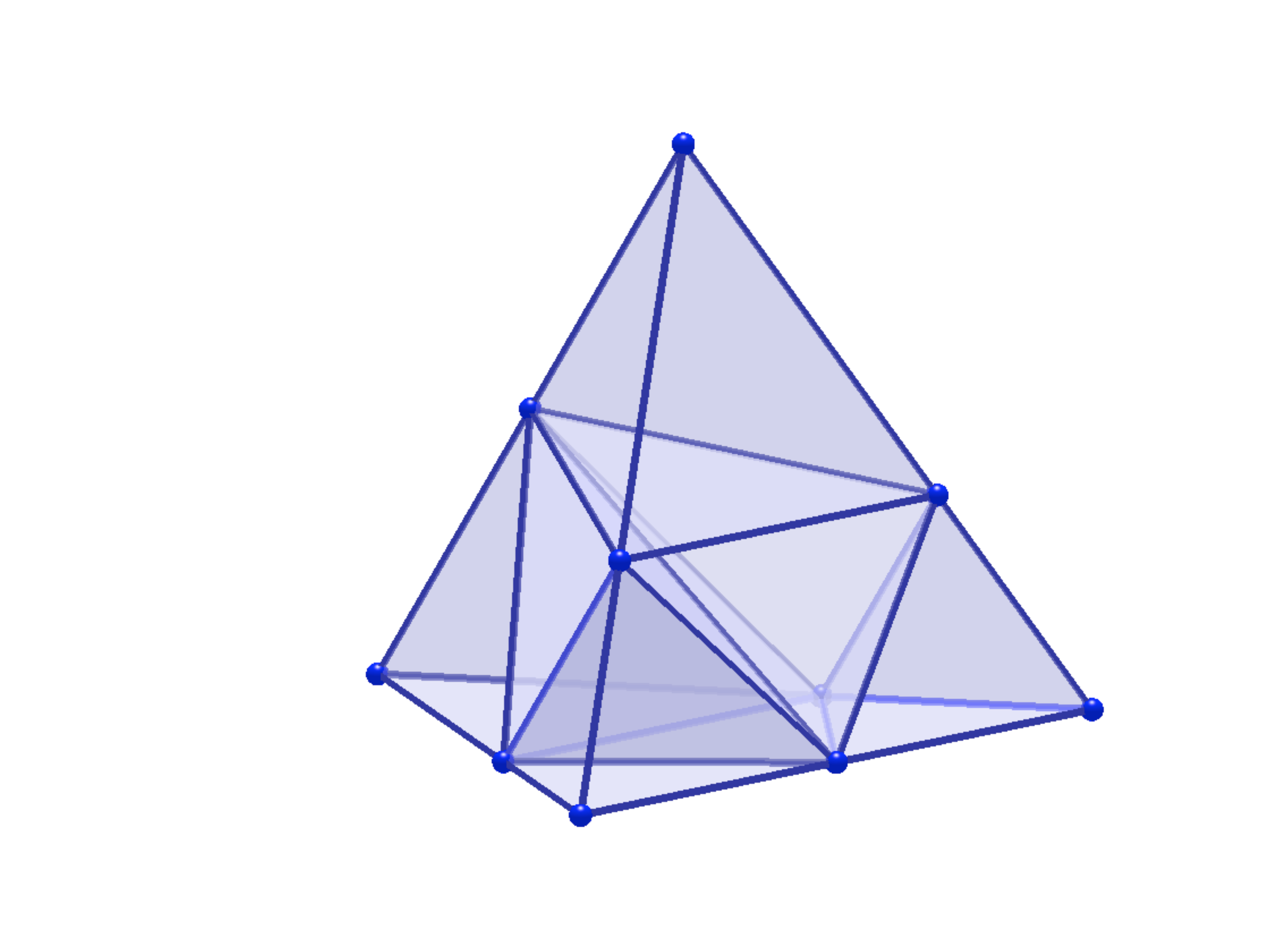}
\end{minipage}%
\hspace{10pt}%
\begin{minipage}[c]{200pt}
\centering
\begin{tabular}{|l|l|l|}
\hline
\multicolumn{3}{|c|}{Quality Statistics}\\
\hline
\multicolumn{1}{|c|}{Quantity} & \multicolumn{1}{c|}{Min}
  & \multicolumn{1}{c|}{Max}\\
\hline
\ $h/R$ & \ \ $0.0345$ & \ \ $\phantom{0}0.712$\\
\hline
\ Face Angle & \ \ $38.87$\textdegree\ \ 
  & \ \ $\phantom{0}86.76$\textdegree\ \ \\
\hline
\ Dihedral Angle\ \ & \ \ $38.44$\textdegree & \ \ $121.37$\textdegree\\
\hline
\ $R/\ell$ & \ \ $0.702$ & \ \ $\phantom{0}0.934$\\
\hline
\end{tabular}\\
\end{minipage}
\caption{A simple subdivision of the regular tetrahedron
  into eight completely well-centered tetrahedra}
\label{fig:regTetSbdv_8a}
\end{figure}

\begin{figure}
\begin{minipage}[c]{160pt}
\centering
\includegraphics[width = 153pt, trim = 322pt 108pt 146pt 115pt, clip]%
  {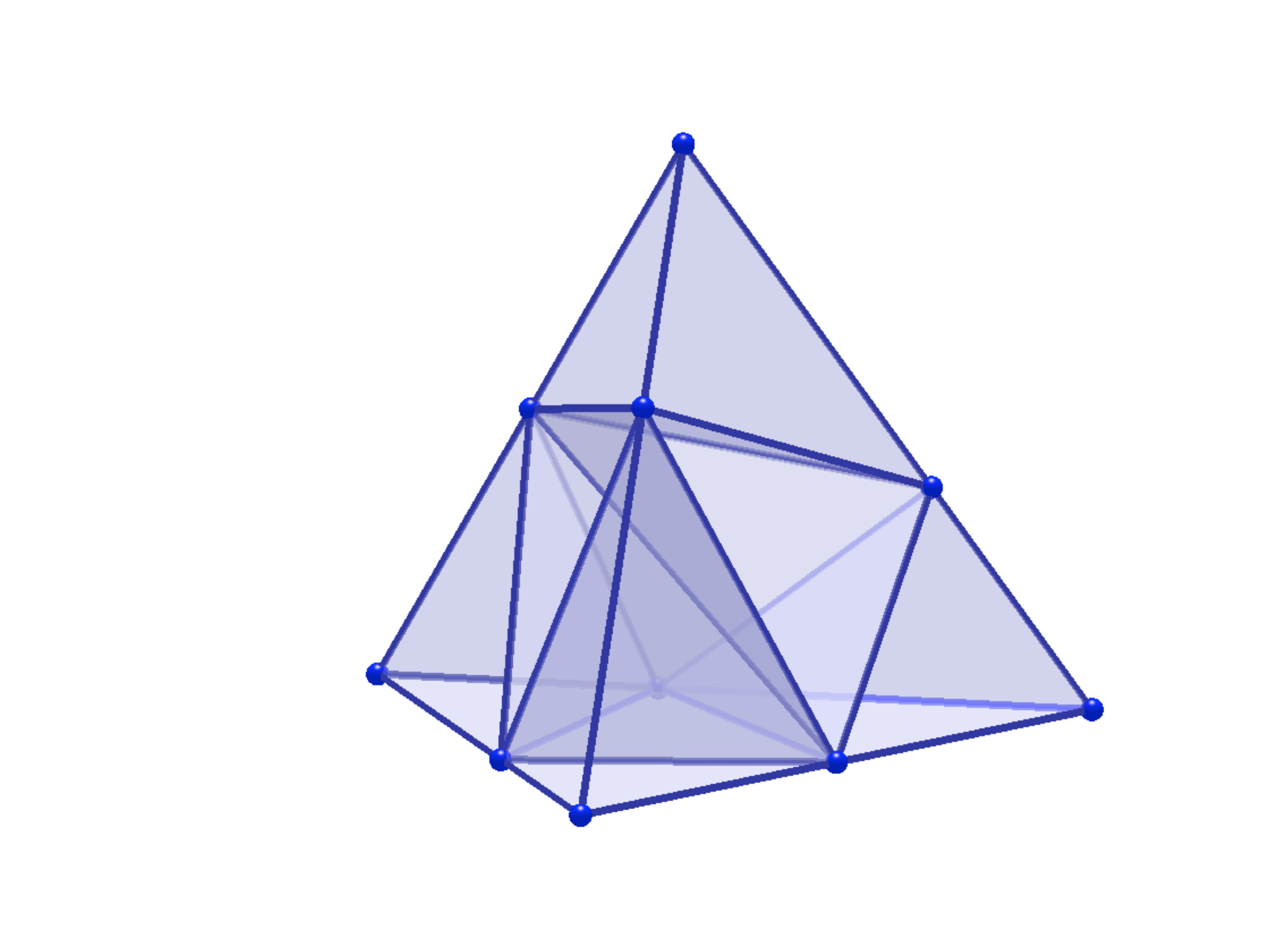}
\end{minipage}%
\hspace{10pt}%
\begin{minipage}[c]{200pt}
\centering
\begin{tabular}{|l|l|l|}
\hline
\multicolumn{3}{|c|}{Quality Statistics}\\
\hline
\multicolumn{1}{|c|}{Quantity} & \multicolumn{1}{c|}{Min}
  & \multicolumn{1}{c|}{Max}\\
\hline
\ $h/R$ & \ \ $0.0448$ & \ \ $\phantom{0}0.584$\\
\hline
\ Face Angle & \ \ $39.63$\textdegree\ \ 
  & \ \ $\phantom{0}87.43$\textdegree\ \ \\
\hline
\ Dihedral Angle\ \ & \ \ $46.72$\textdegree & \ \ $105.95$\textdegree\\
\hline
\ $R/\ell$ & \ \ $0.777$ & \ \ $\phantom{0}0.826$\\
\hline
\end{tabular}\\
\end{minipage}
\caption{Another simple subdividision of the regular tetrahedron
  into eight completely well-centered tetrahedra}
\label{fig:regTetSbdv_8b}
\end{figure}

We can turn this subdivision of the regular tetrahedron
into a well-centered subdivision, however.  By sliding some of
the new vertices along the edges of the regular tetrahedron,
moving them away from the edge midpoints, we can make the
four center tetrahedra well-centered without degrading the
four corner tetrahedra to the point of not being well-centered.
Figures~\ref{fig:regTetSbdv_8a} and~\ref{fig:regTetSbdv_8b}
illustrate two different successful ways we can slide the vertices
along edges of the tetrahedra.
In both cases, the
midpoint vertices that are adjacent to the central edge remain
stationary, to keep the central edge as short as possible.
In Fig.~\ref{fig:regTetSbdv_8a}, the four free midpoints
all slide towards the same edge of the regular tetrahedron.
In Fig.~\ref{fig:regTetSbdv_8b}, the free midpoints
slide along a directed four-cycle through the vertices of the
regular tetrahedron.

\begin{figure}
\centering
\includegraphics[width=320pt, trim=30pt 10pt 0pt 0pt, clip]
  {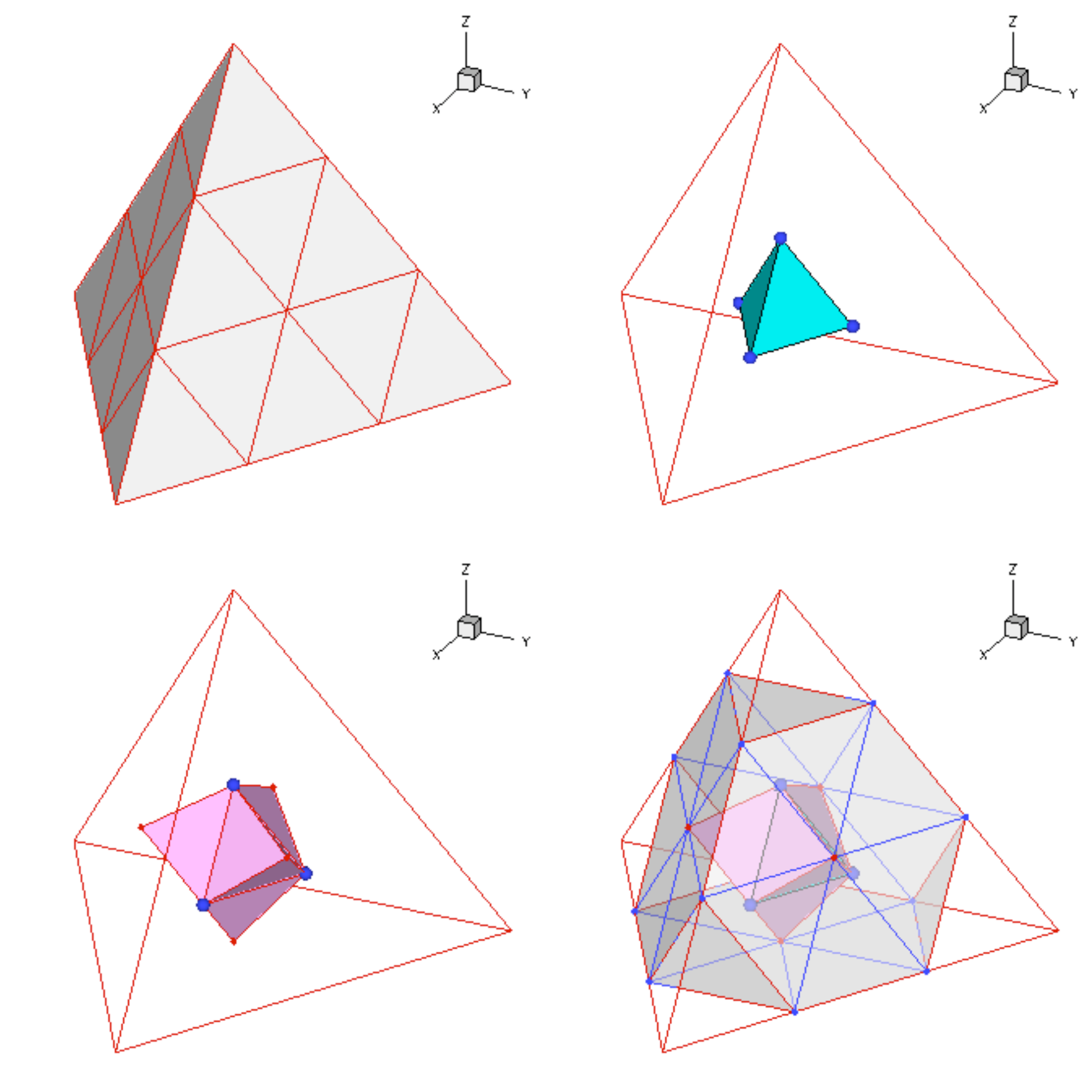}
\caption{A subdivision of the regular tetrahedron that can
  be made completely well-centered through optimization}
\label{fig:tetsubdiv_49_initial}
\end{figure}

\begin{figure}
\centering
\begin{minipage}[c]{165pt}
\centering
\includegraphics[width = 153pt, trim = 60pt 40pt 10pt 20pt, clip]%
  {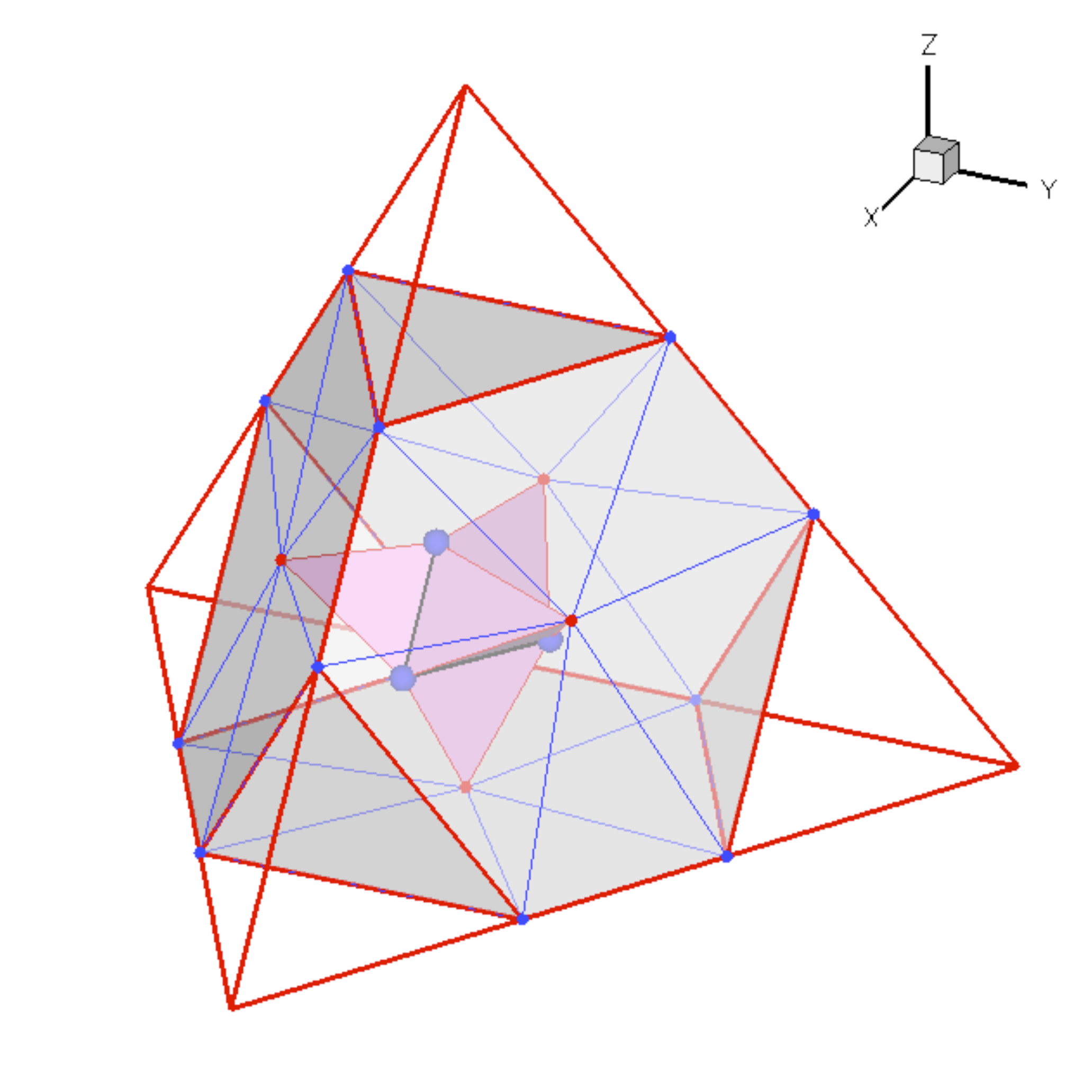}
\end{minipage}%
\hspace{5pt}%
\begin{minipage}[c]{200pt}
\centering
\begin{tabular}{|l|l|l|}
\hline
\multicolumn{3}{|c|}{Quality Statistics}\\
\hline
\multicolumn{1}{|c|}{Quantity} & \multicolumn{1}{c|}{Min}
  & \multicolumn{1}{c|}{Max}\\
\hline
\ $h/R$ & \ \ $0.0146$ & \ \ $\phantom{0}0.845$\\
\hline
\ Face Angle & \ \ $23.36$\textdegree\ \ 
  & \ \ $\phantom{0}89.07$\textdegree\ \ \\
\hline
\ Dihedral Angle\ \ & \ \ $29.93$\textdegree & \ \ $107.73$\textdegree\\
\hline
\ $R/\ell$ & \ \ $0.612$ & \ \ $\phantom{0}1.305$\\
\hline
\end{tabular}\\
\end{minipage}
\caption{The completely well-centered subdivision of the
  regular tetrahedron that results from optimizing the mesh
  shown in Fig.~\ref{fig:tetsubdiv_49_initial}}
\label{fig:tetsubdiv_49_cwc}
\end{figure}

There are also more complicated ways to divide the regular
tetrahedron into smaller well-centered tetrahedra.
Figure~\ref{fig:tetsubdiv_49_initial} shows the basic
structure of a subdivision of the regular tetrahedron
into 49 tetrahedra.  A smaller regular tetrahedron is
placed in the center of the large tetrahedron with the same
orientation as the original.  Each face of the smaller
tetrahedron is connected to the center of a face of the
larger tetrahedron.  At each corner of the
large tetrahedron, a small regular tetrahedron is cut off
of the corner, and the resulting face is connected to a vertex
of the central regular tetrahedron.  After filling in a few more tetrahedral
faces, six more edges need to be added to subdivide octahedral
gaps into tetrahedra.  The mesh shown in Fig.~\ref{fig:tetsubdiv_49_cwc}
is the completely well-centered mesh that results from optimizing the
mesh shown in Fig.~\ref{fig:tetsubdiv_49_initial}.  This subdivision
of the regular tetrahedron is interesting partly because all of
the surface triangulations match and have three-fold radial symmetry.
It is also possible that this type of subdivision will be easier
to use in mesh refinement than the other two subdivisions.

It is not clear whether these constructions can be extended in some
way to create a well-centered subdivision of any well-centered
tetrahedron.  The constructions cannot be extended to create
well-centered subdivisions of all tetrahedra, since both constructions
cut off the corners of the tetrahedron to create smaller tetrahedra
that are nearly similar to the original tetrahedron.  In particular,
the cube corner tetrahedron, i.e., some scaled, rotated, translated
version of the tetrahedron with vertices $(0, 0, 0)$, $(1, 0, 0)$,
$(0, 1, 0)$, and $(0, 0, 1)$, cannot be subdivided into well-centered
tetrahedra in this fashion; one can show that no tetrahedron with
three mutually orthogonal faces is $3$-well-centered
\cite{VaHiGuZh2008}.  Subdividing the cube corner tetrahedron is
particularly interesting, though, because it provides some guidance
regarding what is needed to mesh a cube or, for that matter, any
object having three mutually orthogonal faces that meet at a point.

In fact, the smallest known subdivision of the cube into
well-centered tetrahedra is based on a subdivision of the
cube corner into well-centered tetrahedra.  The picture in
the bottom right corner of Fig.~\ref{fig:cube_194_hsbcwc}
shows a triangulation of a region that fits into the corner
of a cube.  The three visible interior vertices in
Fig.~\ref{fig:cube_194_hsbcwc} are not coplanar with
the cube diagonals, but they are nearly so, and there
is a completely well-centered mesh of the cube corner
that is combinatorially the same as that mesh.  We
actually obtained the mesh of the cube corner tetrahedron
first, and the well-centered mesh of the cube in
Fig.~\ref{fig:cube_194_hsbcwc} was obtained by
replicating the cube corner mesh as described earlier,
adding a vertex at the cube center, computing the Delaunay
triangulation, and optimizing
with the software discussed in \cite{VaHiGuRa2008}.

\section{Conclusions and Questions}

We have discussed some of the properties of well-centered
tetrahedra and seen that it is possible to triangulate
a variety of basic three-dimensional shapes with completely
well-centered tetrahedra.  The triangulations discussed
suggest that it might soon be practical to mesh simple domains in
$\RR^3$ with well-centered tetrahedra.  They also show that
there are a rich variety of well-centered tetrahedra.

The authors hope that it will be possible to build robust
software for meshing three-dimensional domains with
this variety of well-centered tetrahedra, but there is still
significant work to be done before that goal can be reached.
Part of this work is to determine what properties the neighborhood
of a vertex in a $3$-dimensional triangulation must have in order
to permit a well-centered triangulation.  The question is
partially answered in \cite{VaHiGuZh2008}, but a complete answer
is lacking.  The ability to subdivide any 
tetrahedron (or even just any well-centered tetrahedon)
into smaller well-centered tetrahedra would also
be a significant advance toward this goal.

This work also raises some questions that are of more theoretical
interest, though not without practical application.  It
would be interesting to construct the smallest possible
well-centered mesh of the cube.  How might one improve the
lower bounds on the number of tetrahedra needed in a
well-centered mesh of the cube?  Also it is still an open question
whether there are other tetrahedra for which copies of a single
tetrahedron meet face to face and fill space.  Could there be
other well-centered tetrahedra that tile space?  Are there
other families of high-quality tetrahedra that tile space?

\section*{Acknowledgment}
The authors thank Edgar Ramos and Vadim Zharnitsky for useful
discussions.  The work of Evan VanderZee was supported by a fellowship
from the Computational Science and Engineering Program, and the
Applied Mathematics Program of the University of Illinois at
Urbana-Champaign. The work of Anil N. Hirani was supported in part by
an NSF CAREER Award (Grant No. DMS-0645604) and the work of Damrong
Guoy was supported by CSAR (Center for Simulation of Advanced
Rockets).  The CSAR research program is supported by the US Department
of Energy through the University of California under subcontract
B523819.

\bibliographystyle{acmurldoi}
\bibliography{wct}

\end{document}